\documentclass[prb,preprintnumbers,amsmath,amssymb,floatfix,aps,nofootinbib,superscriptaddress,twocolumn,
byrevtex,
    floatfix,
    nofootinbib,
    longbibliography,
    linenumbers]{revtex4}

\usepackage{graphicx}
\usepackage{hyperref}
\usepackage{xcolor}

\usepackage{afterpage}

\newcommand*{\citen}[1]{%
  \begingroup
    \romannumeral-`\x 
    \setcitestyle{numbers}%
    \cite{#1}%
  \endgroup   
}

\usepackage[title]{appendix}
\usepackage{tikz} 
\usepackage{amsmath}
\usepackage{hyperref}
\usepackage[]{subfigure}
\usepackage{float}

\hypersetup{colorlinks,linkcolor={blue},citecolor={blue},urlcolor={cyan}}

\begin{document}
\title{Ca$_{3}$Ru$_{2}$O$_{7}$: Interplay among degrees of freedom and the role of the exchange-correlation}

\author{A.M León}
\affiliation{Facultad de F\'{i}sica, Pontificia Universidad Cat\'{o}lica de Chile, Santiago, Chile.
}
\affiliation{Max Planck Institute for Chemical Physics of 
Solids, Dresden, Germany.}
\affiliation{Faculty for Chemistry, TU Dresden, Bergstrasse 66c, 01069 Dresden, Germany.
}
\author{J. W. Gonz\'alez}
\affiliation{Departamento de F\'{i}sica, Universidad 
T\'{e}cnica Federico Santa Mar\'{i}a, Casilla Postal 
110V, Valpara\'{i}so, Chile.}

\author{H. Rosner}
\affiliation{Max Planck Institute for Chemical Physics of 
Solids, Dresden, Germany.}

\begin{abstract}
Ca$_{3}$Ru$_{2}$O$_{7}$ is a fascinating material that displays physical properties governed by spin-orbit interactions and structural distortions, showing a wide range of remarkable electronic phenomena. Here, we present a density-functional-based analysis of the interplay among degrees of freedom, such as magnetism, Coulomb repulsion (Hubbard-U), and structural degrees of freedom, considering two exchange-correlation methods: Local-Density Approximation (LDA) and Perdew-Burke-Ernzerhof revised for solids (PBEsol). Our goal is twofold: first, to provide a brief overview of the current state of the art on this compound underpinning to the last proposed theoretical models and experimental research, and second, to give another view to model the electronic properties compared with the previous theoretical models. Our findings show that Ca$_{3}$Ru$_{2}$O$_{7}$ displays several electronic states (metal, semimetal, and narrow insulator) as a function of Hubbard-U while it exhibits structural transition depending on the functional. We disentangle the effect of the different degrees of freedom involved, clarifying the role of exchange-correlation in the observed electronic and structural transitions.


\end{abstract}

\maketitle
\section{Introduction}

\footnotetext[1]{Present address: Pontificia Universidad Católica de Chile; andrea.leon@uc.cl }

Physics driven by spin-orbit interactions is among the most important topics in contemporary condensed matter physics\cite{soumyanarayanan2016emergent,witczak2014correlated}. Since the 
spin-orbit interaction is comparable to the on-site Coulomb and other 
relevant interactions create a unique balance between competing interactions that drive complex behaviors and exotic states. In particular, compounds involving transition metals are technologically relevant. However, with their characteristic partially-filled $d$ orbitals, they exemplify this non-trivial competition between interactions of different natures
\cite{kosmider2013large,rejali2020complete,gonzalez2017complex}.
The series Ca$_{n+1}$Ru$_{n}$O$_{3n+1}$ compounds exemplify these
characteristics; these $4d$-electrons systems present a perovskite structure, where the relative rotation and tilting of RuO$_{6}$ octahedra often affect electronic properties depending on the number of layers ($n$ in Ca$_{n+1}$Ru$_{n}$O$_{3n+1}$). Among these, the Ca$_{3}$Ru$_{2}$O$_{7}$ compound (CRO) presents several quantum phenomena such as electronic phase transitions, colossal magneto-resistance, spin density waves, or quantum 
oscillations \cite{cao1997observation, colossal-Yoshida2004, 
kikugawa2007ca3ru2o7,spindensityCa3}. Early works suggested that CRO is a hallmark material exhibiting both metallic and insulator states.
Due to these characteristics, CRO is cataloged as a peculiar 
material\cite{caofrontiers} whose further comprehension could provide new routes to understand the complex interplay among spin, orbital, charge, and lattice degrees of freedom present in $4d$-oxides \cite{markovic2020electronically, xing2018existence}.

\begin{figure}[hb]
\includegraphics[width=0.5\textwidth]{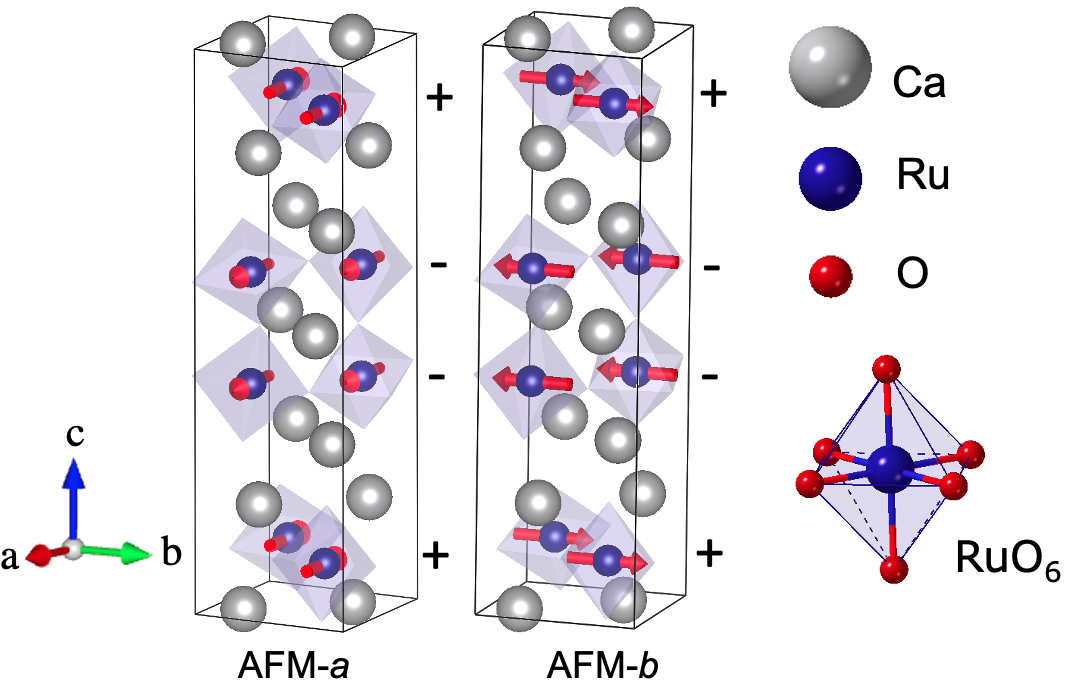}
\caption{Ca$_{3}$Ru$_{2}$O$_{7}$ in two antiferromagnetic phases, AFM-$a$ and AFM-$b$, with ferromagnetic coupling within the layers and antiferromagnetic coupling between the bilayers.
The +/- signs represent the relative orientation of the Ru in-plane magnetic moments: ``$+$" for magnetic moments oriented in the axis direction and ``$-$" for those oriented in the opposite direction.}
\label{fig00}
\end{figure}

Currently, CRO is recognized as a polar metal holding a strong interplay between spin-orbit coupling (SOC) and structural inversion symmetry\cite{markovic2020electronically,horio2021electronic}, which leads to their exotic states. Recent experimental research has given new perspectives to manipulate and continue exploring their quantum states. In this context, have been observed metamagnetic phases \cite{sokolov2019metamagnetic}, which may host mixed textures similar to the chiral magnetic skyrmions, being this an ideal playground to study diverse antiferromagnetic-ferromagnetic states near metamagnetic transitions. Also, it was evidenced by the emergence of polar domains that may be controlled by ferroelectric switching. This last discovery opens new exciting questions about the mechanism behind supporting such domains in polar metals\cite{danilo-2}.
Moreover, it was shown that CRO exhibits extraordinary lattice flexibility because its electronic states can be tuned by manipulating their crystalline axes via external magnetic fields and under pressure\cite{colossal-Yoshida2004,Lattice2021}. The studies mentioned above open a new research stage on the CRO system. The appearance of magnetic textures and polar domains, and the facility for manipulating electronic states under lattice deformation mechanisms, could give novel perspectives to search for the next generation of ferroelectric compounds with novel functionalities \cite{haldiwi2017}.

The CRO structure is defined by RuO$_6$ planes intercalated with 
calcium atoms. Due to the strong spin-orbit coupling present in the Ru
atoms, non-collinear magnetism is expected\cite{johansson2010spin,srnec2008effect}. The interaction between Ru planes depends on the large coupled rotations and tilts of the RuO$_{6}$ octahedra that make up the perovskite-like building blocks of this structure, generating a non-centrosymmetric crystal structure. 
In particular, one of the most striking characteristics of CRO is its phase transitions under temperature, exhibiting a first magnetic transition at the N\'eel temperature of 56 K (T$_{N}$). This first transition is associated with the spins aligning ferromagnetically within each bilayer in the [100] direction but antiferromagnetically coupled between bilayers (AFM-$a$). A second structural and magnetic transition happens at 48 K (T$_{S}$),  where occurs a c-axis lattice parameter compression and the spins are reoriented from [100] to [010] direction (AFM-$b$) \cite{cao1997observation,kikugawa2007ca3ru2o7,yoshida2005}. 

Below T$_{S}$, the in-plane resistivity starts to increase with cooling, showing a local maximum at T $\approx$ 30 K. In early works\cite{cao1997observation}, the phase transition at T$_{s}$ has often been discussed as a metal-insulator transition, and the temperature region between 30 K and T$_{S}$ as an insulating phase of Ca$_{3}$Ru$_{2}$O$_{7}$; the understanding of the nature of these phases (either metallic or insulator) has been controversial in the literature\cite{xing2018existence}. In this context, it was suggested that different growth batches of Ca$_{3}$Ru$_{2}$O$_{7}$ crystals yield differing low temperature properties, ranging all the way from the metallic ground states \cite{spindensityCa3,kikugawa2007ca3ru2o7,baumberger2006nested,lee2007pseudogap} to insulating phases\cite{cao1997observation,cao2003quantum,karpus2006spectroscopic}. Conciliating the different experimental observations have been an intense topic of research and debate for more than one decade\cite{spindensityCa3,colossal-Yoshida2004,baumberger2006nested,kikugawa2007ca3ru2o7,lee2007pseudogap,cao1997observation,cao2003quantum,karpus2006spectroscopic}.

Using a theoretical approach based on the density functional theory (DFT), we explore the electronic and magnetic properties of CRO crystals to evidence the conditions that lead to either metallic or non-metallic ground state solutions. Here, we explore the interplay among Coulomb repulsion (U-Hubbard), spin-orbit effects (SOC), structural degrees of freedom (positions and volume), and exchange-correlation. We compare our results against the latest experimental and theoretical results on CRO. 
%
Furthermore, we contrast the electronic band structure of CRO using two exchange correlations: PBEsol (PS) and LDA approximations and performing a linear response approximation calculation implemented by Cococcioni-Gironcoli \cite{cococcioni2005linear}. We establish the upper and lower limits of the optimal U-Hubbard values. The contrast of our results against the experimental evidence reveals that LDA+U ($0.2 \leq U \leq 2.0$ eV) models properly the electronic and structural properties of CRO.


\section{Theoretical advances\label{sec:adv}}
As we mentioned before, one of the biggest problems was to conciliate the metallic or insulator nature of the phase between 30 $ \lessapprox$ T $<$ T$_{S}$, in which some experimental works suggested a metal/semi-metal behavior\cite{spindensityCa3,kikugawa2007ca3ru2o7,baumberger2006nested,lee2007pseudogap} and others an insulator phase \cite{cao1997observation,cao2003quantum,karpus2006spectroscopic}. From a theoretical point of view, early DFT studies have reported electronic properties under different approximations. Collinear calculations predict AFM ground state with an FM and AFM spin coupling intra- and inter-layer\cite{singh2006electronic}, respectively, agreeing with the experimental report\cite{yoshida2005}.
Along with the FM configuration, it was evidenced that the system exhibits a nearly half-metallic state.
Later, noncollinear calculations explored the role of U and SOC on the electronic properties, indicating the U region in which the system could be metallic or insulator, which in turn depends on the magnetic moment orientation\cite{Liu-cro-327}.

Recent experimental and theoretical studies have given new perspectives about the nature of CRO below T$_{S}$, reaching a better agreement between theory and experiments. It was revealed that CRO does not have a real insulator phase
since it was evidenced a pseudogap behavior around $\Gamma$ point and several bands crossing the Fermi level \cite{markovic2020electronically,horio2021electronic}. Despite this advance, understanding the mechanism that causes the phase transitions around T$_{S}$ is still controversial. On the one hand, it is proposed a Lifshitz-like transition at T$_{s}$ together with a new structural transition at T $ \approx$ 30 K as a result of the RuO$_{6}$ octahedral distortions and hybridization changes induced by the enhancement of the Coulomb interactions upon cooling \cite{danilo}. On the other hand, it is suggested that the phase transition at T$_{s}$ arises due to the thermal population reduction, opening a gap at the Fermi level. These hybridization changes induce an electronic energy gain favoring the spin reorientation through Rashba-type spin-orbit coupling, accompanied by structural changes simultaneously \cite{markovic2020electronically}. Previous theoretical works have discussed the role of fundamental interactions such as Coulomb and spin-orbit coupling considering GGA\cite{Liu-cro-327,danilo} and PS\cite{danilo} approximations. However, in this work, we aim to gain insight into the interplay among several degrees of freedom, such as magnetism, SOC, Hubbard-U repulsion, and structural degrees of freedom, considering the role of the exchange-correlation (LDA and PS) in the electronic and structural properties of CRO.\\

Why is it so complicated to understand the electronic structure of CRO? From a theoretical point of view, studies based on DFT have observed that the unconventional magnetic and electronic properties are dictated by the competition of multiple degrees of freedom (charge, spin, SOC, lattice) that makes the electronic states highly sensitive to the electronic interactions and exchange of correlation. Early studies using GGA approximation evidenced a metallic state and highly anisotropic behavior of CRO. They suggested for the first time the crucial role of spin-orbit interaction, whereas a metal-insulator transition happens by the interplay of SOC and Coulomb repulsion (U) (for U $>$ 3 eV)\cite{Liu-cro-327}. Later, other studies considering spin-orbit interaction with LDA and GGA evidenced that the system has a semimetal state. Here, it was evidenced that GGA fails to describe the band structure at $\Gamma$ point through comparing with ARPES spectra \cite{markovic2020electronically}. Finally, recent work has performed studies using GGA along with PS approximation and considering the interplay among SOC+U+structural degrees of freedom; this study revealed that for small U values, it is possible to control the band occupation around the Fermi level, in which for U $\geq$ 1.4 give good agreement with the ARPES spectra \cite{danilo}. However, it is necessary to propose a new structural phase transition not reported by experiments so far.



\section{Calculation Details}

We perform a theoretical analysis using the density functional theory 
with spin-orbit coupling (SOC). We use the plane-wave pseudopotential 
method implemented in the Vienna ab-initio simulation package 
(VASP)\cite{Kresse1996} within the PBEsol functional (PS)\cite{Perdew2008} 
and the Local-density approximations (LDA)\cite{ceperley1980ground}. 
PS belongs to generalized gradient approximations (GGA), 
and it gives structural properties closer to the experimental values 
when compared to the results of PBE\cite{dongho2015density}. The electronic valence considered are: Ru: 5$s^{1}$4$d^{7}$ and O: 2$s^{2}$2$p^{4}$ (for LDA and PS), and Ca: 3$s$3$p$4$s$/Ca: 2$s$$p^{6}$$s^{2}$$d^{0.01}$, for PS/LDA respectively.
We use a plane-wave energy cutoff of 650 eV and set a  
Regular Monkhorst-Pack grid of 5$\times$5$\times$3 to perform the atomic relaxation and  7$\times$7$\times$5 to perform the self-consistent calculation. We use a fine k-grid 14$\times$14$\times$5 within the tetrahedron method for the density of states.
We perform the structural optimization of the unit cell until a force convergence threshold of at least 10$^{-3}$ eV/\AA\ per atom.

To consider the electronic correlation effects in d-orbitals of Ru atoms, we consider a range of Hubbard on-site Coulomb parameters through the Dudarev approximation \cite{dudarev1998electron}. Also, we employ the Liechtenstein scheme for comparison propose \cite{liechtenstein1995density}. Later, we employ the linear response to estimate the optimal Hubbard-U; we follow the linear approach method\cite{cococcioni2005linear}. The linear response was computed by introducing the interacting and non-interacting occupation response with respect to the localized perturbations up to $V$ = $\pm$ 0.2  eV in the Ru atoms; we compare the Ru $d$-orbital occupation upon the perturbation $V$ along LDA/PS, and with/without SOC.



\section{Results and Discussion}

CRO crystallizes in a orthorhombic structure with space group\cite{yoshida2005} Bb2$_{1}$m  (a full description of the structure is available in ref.~\citen{tesis-igor}). 
The atomic positions are taken from the neutron diffraction measurements\cite{yoshida2005} at 8 K. To reproduce the different collinear AFM configurations, the unit cell includes 48 atoms; there are 12 Ca atoms, 8 Ru atoms, and 28 O atoms. Fig. \ref{fig00} shows the lattice structure of CRO, with two possible magnetic configurations experimentally reported at T $<$ 48 K and T $>$ 56 K called AFM-$b$ and AFM-$a$, respectively. The $\pm \bf{a}$ and $\pm \bf{b}$ notation means the ferromagnetic states with the direction in which the in-plane magnetic moments align. Additionally, we introduce the FM-$b$ configuration that means a FM coupling between layers with the magnetic moment aligned along b direction.


\subsubsection{Exchange–correlation effects, U = 0 eV.}

In the PS approximation, the ground state presents an AFM-$b$ configuration with a magnetic moment ($m$) of $1.36$ $\mu_{B}$ (calculated in a Wigner-Seitz (WZ) radius 1.323 \AA). The energy difference between the AFM-$b$ and AFM-$a$ configuration is 0.39 eV (E$_{AFM-a}$ $-$ E$_{AFM-b}$). Within the same approximation, the FM-b phase presents a total magnetic moment of 1.92 $\mu_{B}$; this value overestimates $m$ about a 6\% concerning the $m$ measured by previous experimental reports\cite{yoshida2005}. The energy differences between the AFM-$b$ state and FM-$b$ phase is 1.69 meV/Ru (E$_{FM-b}$ $-$ E$_{AFM-b}$).

In contrast, the LDA approximation reveals an FM-$b$ ground state with a low total magnetic moment of 0.83 $\mu_{B}$. This state is more stable than AFM-$b$ and AFM-$a$ configuration with an energy difference of  -5.85 meV/Ru (E$_{FM-b}$ $-$ E$_{AFM-b}$) and -2.35 meV/Ru (E$_{FM-b}$ $-$ E$_{AFM-a}$), respectively. Besides, the AFM-$a$ phase has lower energy than the AFM-$b$ phase by -2.67 meV/Ru. Despite the mismatch in the magnetic stability order between PS and LDA, both approximations reproduce the experimental crystal structure\cite{yoshida2005} (see table \ref{is2-leng}, calculation at the experimental volume 580.04 \AA$^{3}$, considering positions degrees of freedom).


Going further, we perform a volume cell relaxation; these values are reported in table \ref{S1-vol-is3} of supplementary material (SM), and our results for the most stable phase are shown in table \ref{S2-energy-is3}. We can see that under volume relaxation do not change the most stable magnetic ground state.

\begin{table}[H]
 \centering
    \begin{tabular}{c|c|c|c|c}
\hline
Appr. & Magnetic conf. & m (${\mu_B}$) & V(RuO$_{6}$) (\AA$^3$) & $l$(Ru-O) (\AA) \\
\hline
LDA    & FM-$b$  & 0.83 & 10.53  & 1.992\\
PS    & AFM-$b$  & 1.36 & 10.54  & 1.993  \\
Exp\cite{yoshida2005}   & AFM-$b$ & 1.8 & 10.53  & 1.992  \\
       \hline
    \end{tabular}
    \caption{Magnetic configuration (conf.) ground state,  
    $m$ the magnetic moment per Ru-atom, V(RuO$_{6}$) is the  RuO$_{6}$ octahedral volume (\AA$^{3}$) and  $l$(Ru-O) is the Ru-O bond average length (\AA) for the most stable magnetic configuration considering U $=$ 0 eV. The third-row label as ``Exp" corresponds to the experimental data\cite{yoshida2005}.}
    \label{is2-leng}
\end{table}

\begin{figure*}
{\includegraphics[width=0.45\textwidth]{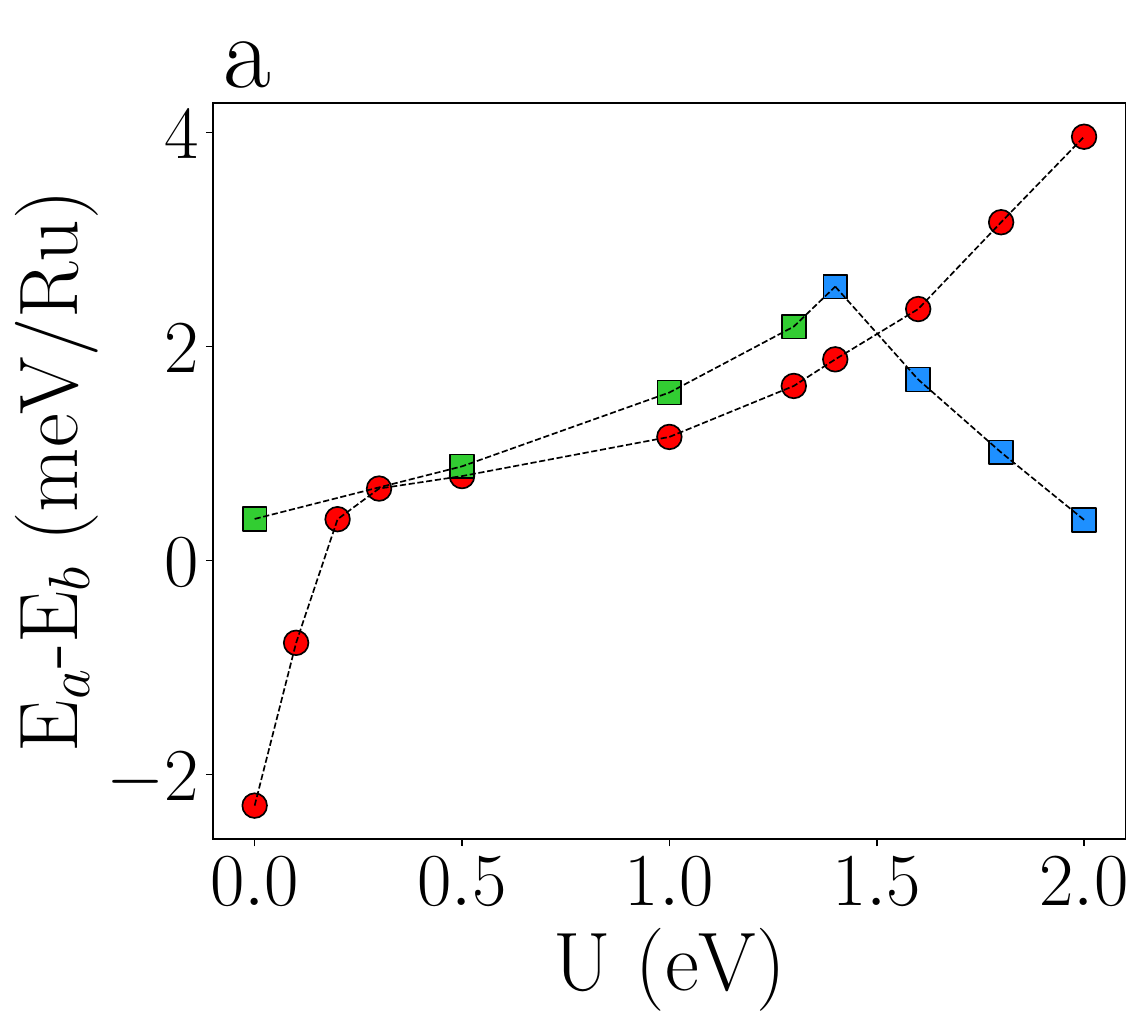}}{\includegraphics[width=0.46\textwidth]{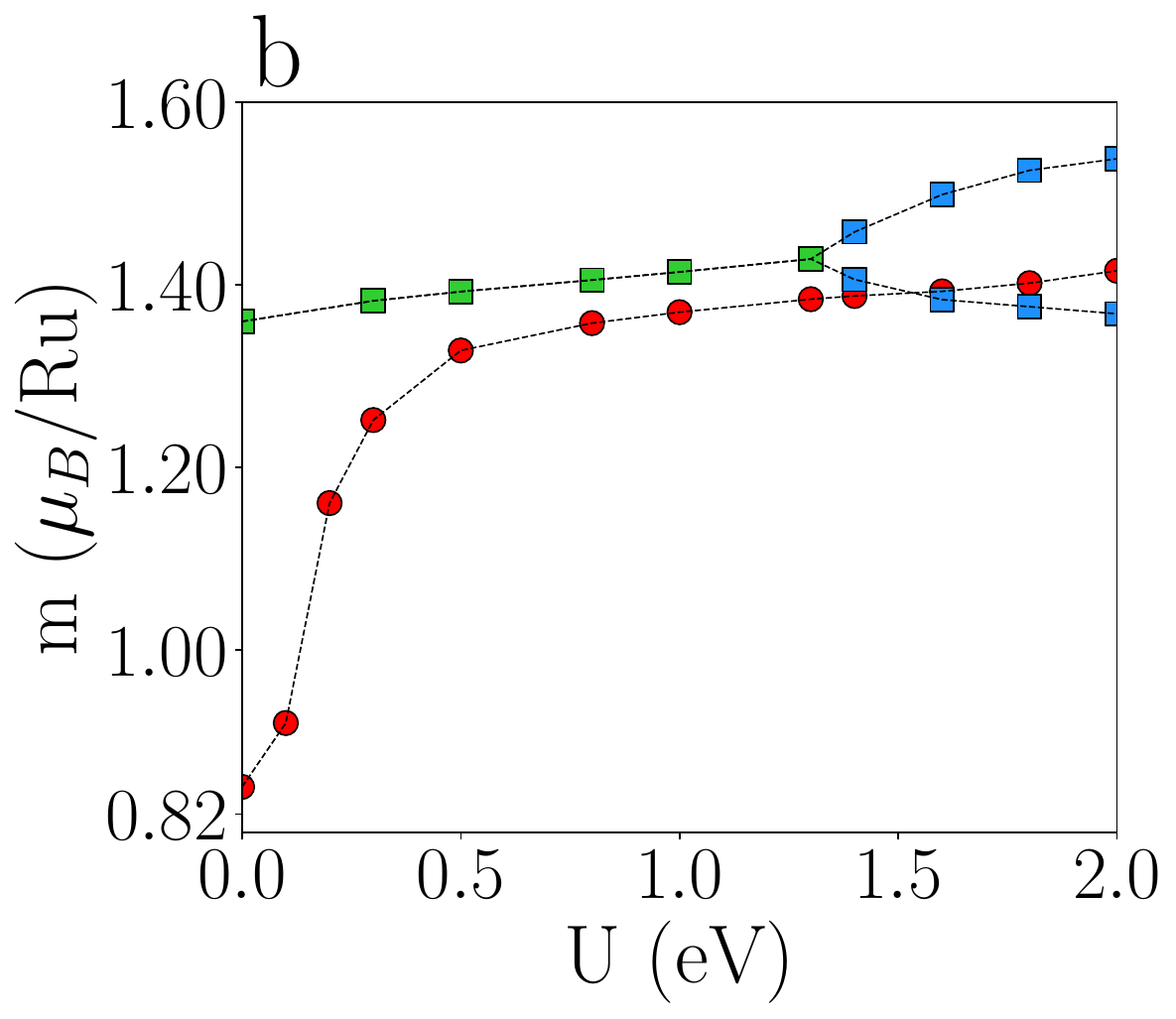}}
\includegraphics[width=0.45\textwidth]{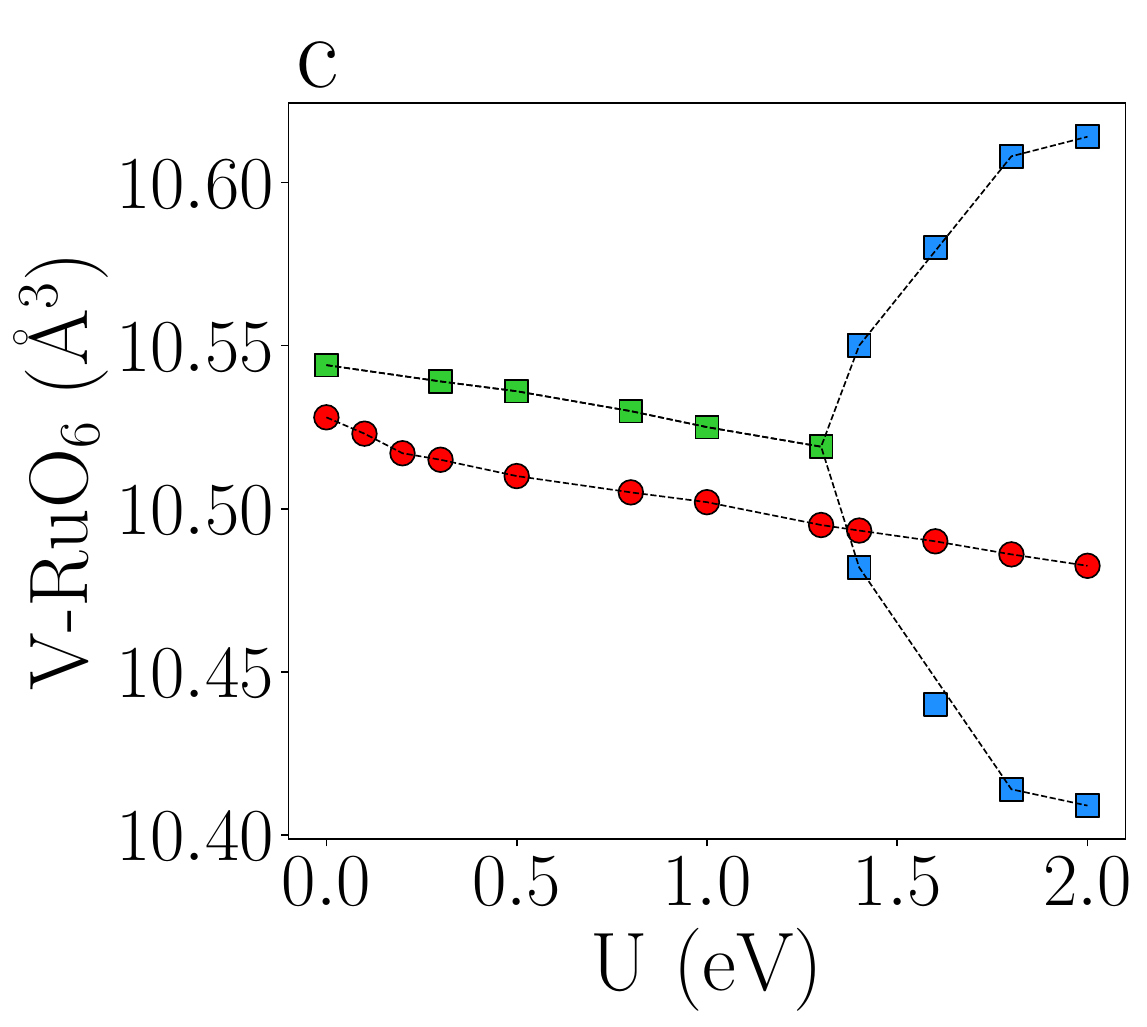}
\hspace{0.4cm}{\includegraphics[width=0.4\textwidth]{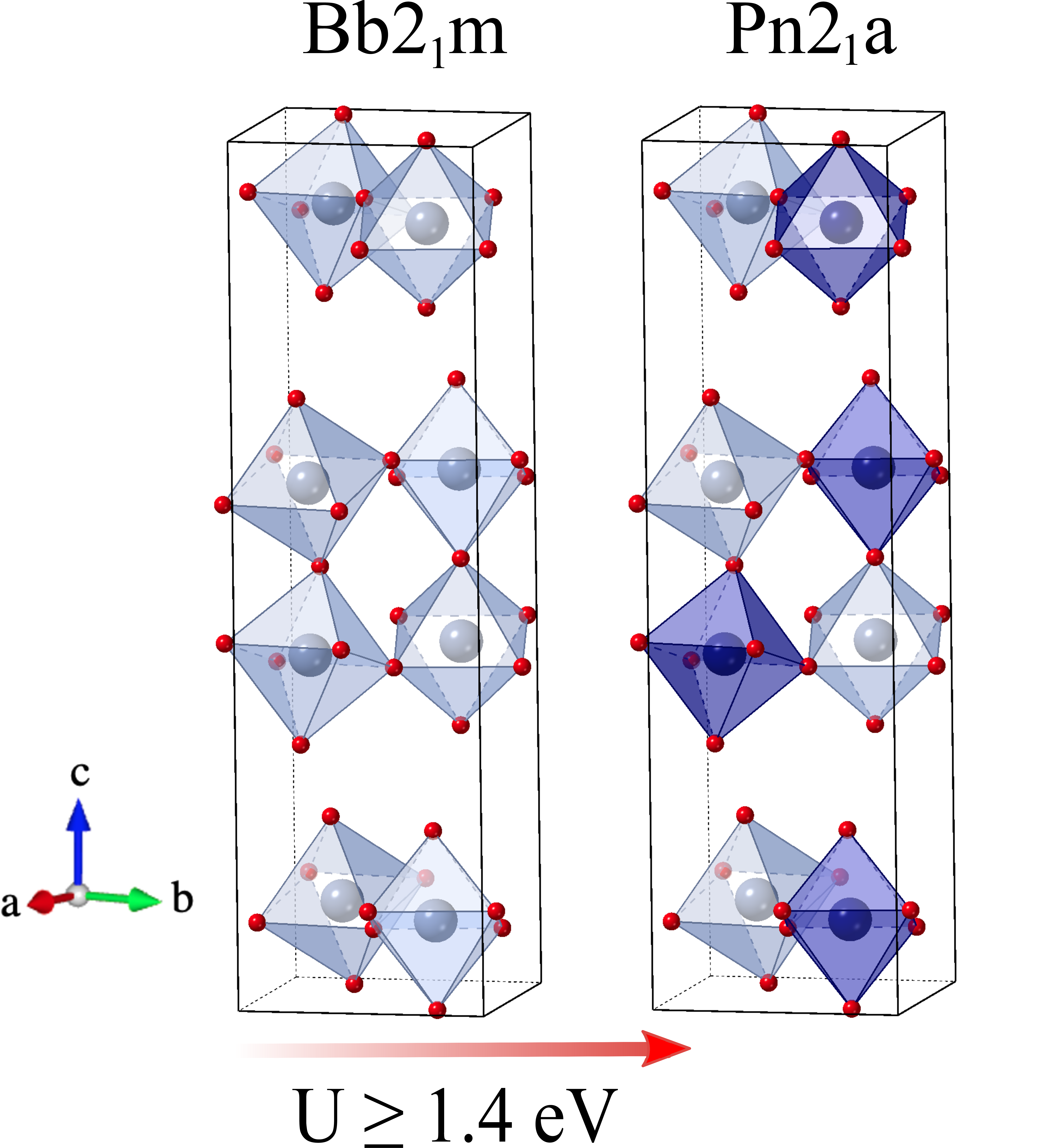}}
\caption{Effect of electron-electron repulsion (Hubbard-U) on the physical properties of CRO. In
(a) magnetic anisotropy E$_{a}$-E$_{b}$, (b) magnetic moment per Ru atom (measured in the Wigner-Seitz radius (1.323 \AA)), and (c) Volume of RuO$_6$ octahedra as U increases from 0 to 2.0 eV. Red and green dots correspond to the crystal phase Bb2$_{1}$a with LDA and PS approximation, respectively, and blue dots correspond to the Pn2$_{1}$a structure with PS approximation. The inset in panel (b)  shows the total magnetic moment  (m$_{T}$) by Ru atom for a ferromagnetic configuration (FM-b), dashed black line indicates the experimental magnetic moment.
Note in panels (b) and (c) for U $\geq$ 1.4 eV and PS approximation, the Ru atoms in the Pn2$_{1}$a phase present two different crystallographic environments and thus two different magnetic moments and volumes. The bottom right panel displays the structural change from Bb2$_{1}$m to Pn2$_{1}$a for U $\geq$ 1.4 eV; the different octahedra color on the Pn2$_{1}$a structure symbolizes the different environments in this cell.}
\label{fig-2}
\end{figure*}


\subsubsection{Electron-electron correlation effects.}
Now we will study the effect of electron-electron repulsion on the electronic and structural properties considering positions degrees of freedom. Based on previous works\cite{danilo}, we use Hubbard-U values in the $\left[0,\,2\right]$ eV range and considering the volume given by the experimental reports\cite{yoshida2005}.
Fig. \ref{fig-2} shows the effect within the PS and LDA approximations of the Hubbard-U interaction for the magnetic anisotropy $\Delta E= E_a-E_b$ (E$_{a}$ $=$ E$_{AFM-a}$ and E$_{b}$ $=$ E$_{AFM-b}$), the magnetic moment per Ru atom $m$, and RuO$_6$ octahedra volume (V-RuO$_6$).

On the one hand, we will start with the LDA analysis, represented with red dots in Fig.  \ref{fig-2}. The magnetic anisotropy as a function of the Hubbard-U is presented in Fig. \ref{fig-2} (a). When the Hubbard-U interaction is included, the AFM-$b$ ground state is restored for U $\geq$ 0.2 eV (this configuration became more stable as $\Delta$E $>$ 0). 
In Fig. \ref{fig-2} (b), as expected, we observe that the U-term increases the magnetic moment rise, from $m =$ 0.84 $\mu_{B}$ at U $=$ 0 to $m =$ 1.16 $\mu_{B}$ at U $=$ 0.2 eV, then the magnetic moment increases monotonously to U $=$ 2.0 eV.  Finally, in Fig. \ref{fig-2} (c), we found that the octahedra volume decreases as the U-term increases.

On the other hand, for the PS approximation represented with blue and green dots in Fig. \ref{fig-2}, we can observe a non-monotonous behavior as the U-term increases. Here, we identify the critical U-term of $U_{c} = 1.4$ eV. When U increases above the $U_{c}$, a spontaneous structural transition appears, changing from Bb2$_{1}$m to Pn2$_{1}$a structure (depicted in the bottom-right panel). This transition could be associated with the interplay of the SOC+U and structural degrees of freedom \cite{danilo}.
In panel (a), after $U_c$ we note a drop in the anisotropy value, where for U = 2 eV we recover the value obtained for U $=$ 0. In panel (b), the magnetic moment after $U_c$ shows two different values associated with the two crystallographic environments observed (see Fig. \ref{fig-2} bottom-right panel); note that the average between the average magnetic moment follows the same trend as before $U_c$. A similar behavior appears when we look at the volume of the RuO$_6$ octahedron in panel (c). Additionally, we have verified that the spontaneous structural symmetry change is independent of the structural degrees of freedom; further comments are in section I of the SM.\\


To properly describe the AFM-$b$ ground state within the LDA approach, it is necessary to consider a large U-Hubbard ($U \geq 0.2$ eV). For $U < 0.2$ eV, the most stable system is the AFM-$a$ configuration (see Fig. \ref{fig-2}(a)). Furthermore, comparing Fig. \ref{fig-2}(a) and (b), we can identify that the regions with AFM-$a$ as the ground state, $E_{a}-E_{b} < 0$,  the magnetic moment is smaller, $m < 1.0 \, \mu_{B}$. In the regions with AFM-$b$ as the ground state, $E_{a}-E_{b} > 0$, the magnetic moment is larger than $1 \, \mu_{B}$.
Since the LDA approximation does not reproduce the experimental magnetization of the ground state for smaller U values ($U< 0.2$ eV), to discuss the theoretical results with the experimental data, we compare the band structure within PS and LDA approximations considering $U_{\mathrm{PS}} = 0$ and $U_{\mathrm{LDA}}=0.5$ eV, respectively.

We should note that the magnetic moment observed for PS and LDA is lower than the magnetic moment experimentally reported (see Table \ref{is2-leng}). This is because the plane-wave approximation yields minor uncertainties in predicting localized quantities, like the magnetic moment and the charge per atomic site. The $m$ displayed in Fig. \ref{fig-2}(b) is measured in the WS radius, then lower than the total magnetic moment. In order to compare our $m$ results with the magnetic moment reported by the experiment for the AFM-b phase ($m$ $\sim$ 1.8 $\mu_{B}$)\cite{yoshida2005}, we compute an FM-b configuration by completeness in order to obtain the total magnetic moment by Ru atom as a U function for both approximations; see details in the SM (see Fig. \ref{m-fm}).



\begin{figure*}[ht]
\includegraphics[width=1.0\textwidth]{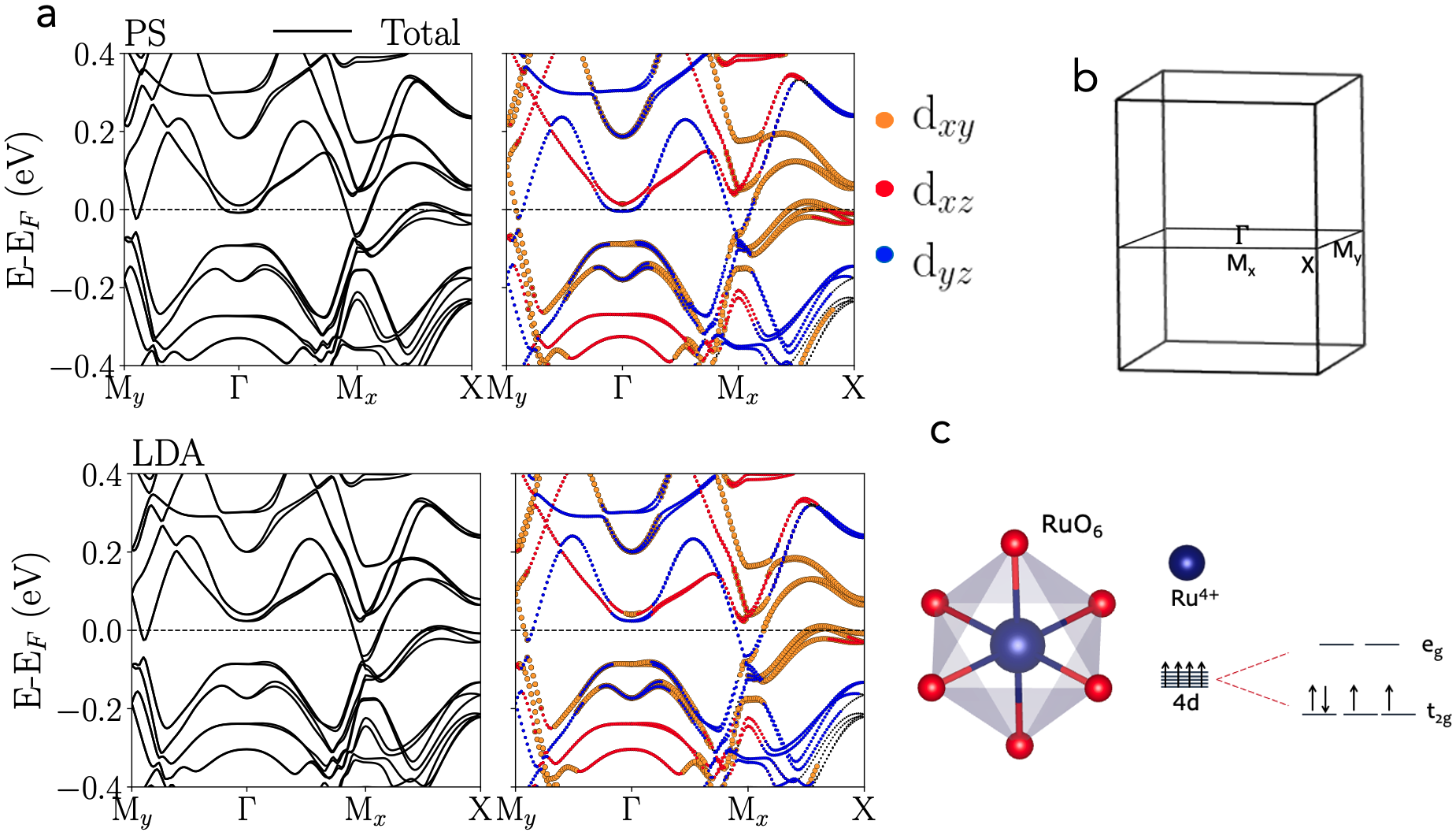}
\caption{(a) The upper and lower panel shows the band structure along PS and LDA calculation, respectively, for U $=$ 0 and 0.5 eV in each case.
Right panel: total bands and the decomposed orbital bands, in which the orbitals have a 25-75\% of occupation per orbital. b) Surface Brillouin zone of Bb2$_{1}$m structure. c) diagram of the crystal field splitting of the Ru 4$d$ states in the octahedral environment. The orbital projection is considered with octahedra aligned to the unit cell; thus, in this reference, the t$_{2g}$ and e$_{g}$ orbitals correspond to (d$_{xy}$,d$_{xz}$,d$_{yz}$) and (d$_{x^{2}-y^{2}}$,d$_{z2}$), respectively.}
\label{fig-bn-u0}
\end{figure*}


The band structure and its projection in atomic orbitals for the PS and LDA approximations are 
shown in Fig. \ref{fig-bn-u0}. 
We will compare systems with similar magnetic anisotropy and magnetic moments (see Fig. \ref{fig-2} a-b); 
for the PS approximation, we will take a value U $=$ 0 eV, and for the LDA approximation, 
we choose a U $=$ 0.5 eV.
In Fig. \ref{fig-bn-u0}, we can see that the band 
structure and its projection are very similar, independent of the approximation used. The system exhibits a 
metallic character with subtle differences at the Fermi level for both approximations, mainly around the $\Gamma$ point.
The valence band is mainly composed of the contribution of atoms of Ru and O (see Fig. \ref{S3-lda-ps-DOS}). The Ru ions are embedded 
in an octahedral environment formed by the oxygen atoms; the octahedral crystal field splits the 
Ru 4$d$ atomic levels, raising the e$_{g}$ and decreasing the t$_{2g}$ orbitals. This configuration yields a low-spin state, with the four 4$d$ electrons partially filling the three t$_{2g}$ orbitals, leaving two unpaired electrons per Ru atom\cite{tesis-igor} aligned as seen
in the right panel of Fig.\ref{fig-bn-u0} and in the partial density of states displayed in SM section (see Fig. \ref{S3-lda-ps-DOS}). The orbital projection is considered with octahedra aligned to the unit cell; thus, in this reference, the t$_{2g}$ and e$_{g}$ orbitals correspond to (d$_{xy}$,d$_{xz}$,d$_{yz}$) and (d$_{x^{2}-y^{2}}$,d$_{z^{2}}$), respectively.


In the case of PS, we can observe a metallic character at $\Gamma$; however, this is not observed in the LDA approximation. We observe a local band gap of 65 meV around the $\Gamma$-point 
in the LDA approximation. 
For both approximations, we identify a Dirac-like behavior at M$_{x}$, in agreement with the ARPES measurements\cite{markovic2020electronically}. Along M$_{x}$-$X$ path, a small band crossing around the Fermi level is found. 
The projected band structure around the Fermi level reveals a dominant contribution (between a 25\% to 75\% per orbital) of the d$_{xy}$, d$_{xz}$, and d$_{yz}$. The d$_{x^{2}-y^{2}}$, d$_{z^{2}}$ orbitals represent less than 25\% of the projection, and therefore they are not displayed. In both projections, we can see that around the $\Gamma$, the dominant states belong to the hybridization between d$_{xy}$-d$_{yz}$ (below the Fermi level) and d$_{yz}$-d$_{xz}$ without hybridization (above of Fermi level). Along M$_{y}$-$\Gamma$ the bands are mainly composed of d$_{xy}$ and d$_{yz}$ orbitals.
The Dirac-like bands are characterized by d$_{xy}$, d$_{yz}$ orbitals, and finally, along M$_{x}$-$X$ path, the most dominant contribution became 
from d$_{xy}$ and d$_{xz}$ states (around the Fermi level). 

In the case of LDA for $U = 0$, its is found a different trend on the band dispersion as well as for AFM-$a$ (ground state) and AFM-$b$ configuration. Here we evidenced a metallic behavior along all the high symmetry points (details in SM, Fig. \ref{S4-lda-bands-A-B}).\\


\begin{figure*}[ht]
\includegraphics[width=1.0\textwidth]{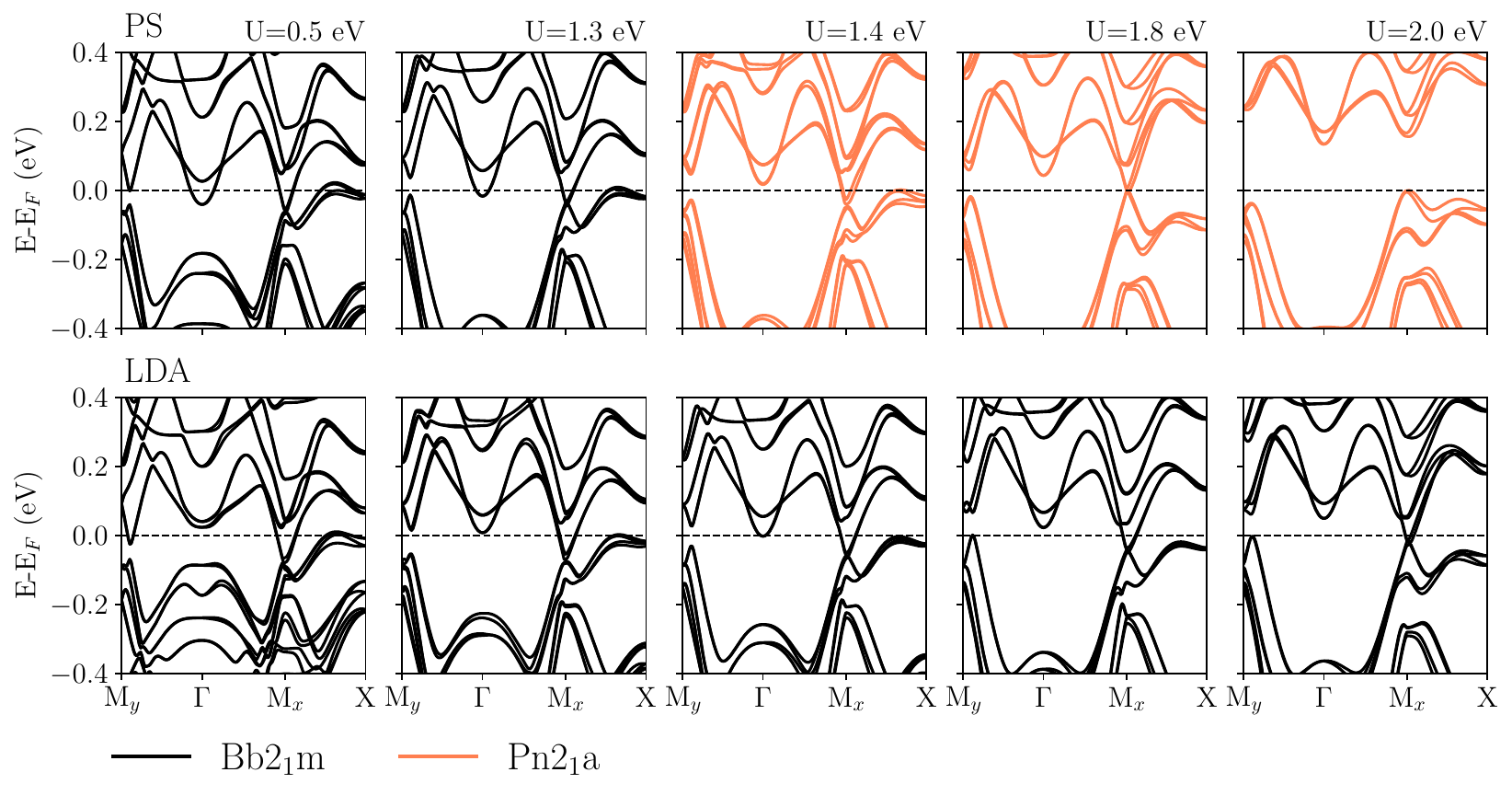}
\caption{The Upper and lower panel displays the band structure computed with  PS and LDA approximation, respectively. The black band corresponds to the Bb2$_{1}$m structure and the color band to the Pn2$_{1}$a phase. Calculation considering SOC+U+ positions degrees of freedom.}
\label{fig-band-u}
\end{figure*}

We systematically study the effect of U-Hubbard repulsion on the band structure; in Fig. \ref{fig-band-u}, we show the evolution of the band structure for PS and LDA approximations as $U$-term changes in a range from $0.5$ eV to $2.0$ eV.
In general, for the same U-term value, there are similarities between PS and LDA; below the Fermi level, the bands around the $\Gamma$ point that are mainly composed by d$_{yz}$ and d$_{xy}$ orbitals, these orbitals are pushed to lower energies as the U-term increases, favoring the hybridization between these orbitals (see Fig. \ref{S-bnd-partial}). 
Along M$_{x}$ to X path, we can observe that these states are slightly pushed below the Fermi level. These bands are rather modified due to they are mainly composed of d$_{xy}$-orbitals being close to integer filling and then exhibiting a less grade of hybridization (see Fig. \ref{fig-bn-u0} right panel). Besides, from M$_{y}$-$\Gamma$ as the Hubbard-U increases, the bands are pushed to higher energies inducing a band distribution aiding the occupation of the d$_{xz}$ orbitals, which slightly hybridize with the d$_{xy}$ orbitals just below the Fermi level (see Fig. \ref{S-bnd-partial}).
Although, for PS approximation, we can distinguish three ranges of $U$-term values. The first region, between 0 $\leq$ U $\leq$ 1.4 eV, is characterized by a metallic state (with a band crossing the Fermi level at $\Gamma$, in the first three plots of the top panel in Fig. \ref{fig-band-u}). 
In the second region, for 1.4 $\leq$ U $\leq$ 1.8 eV, we can observe that the bands at $\Gamma$ and the Dirac-like bands are pushed to high energies given to the system a semimetallic character (due to the gap around $\Gamma$ point). These changes are simultaneous with the phase transition when symmetry changes from Bb2$_{1}$m to Pn2$_{1}$a, as discussed before. 
In the last region, U $\geq$ 1.8 eV, the Dirac-like bands are suppressed due to the orbital occupation rearrangement favoring the 
band hybridization, in which the d$_{yz}$ are shifted to higher energies separated by a gap of 0.16 eV of the hybridized d$_{yz}$-d$_{xz}$ orbitals below the Fermi level (see Fig. \ref{Fig-u20-ps}, for U $=$ 2 eV). The suppression of the Dirac-like bands allows the emergence of a narrow bandgap insulator phase with Pn2$_{1}$a structure along with PS approximation; also, we can observe the gap oppening around the Fermi level in the total density of states displayed in Fig. \ref{S-dos-u}. 

When examining the band structure along the LDA approximation (lower panels of Fig. \ref{fig-band-u}), unlike in the PS case for lower U values (U $<$ 1.4 eV), we do not observe bands crossing the Fermi level in the vicinity of the $\Gamma$ point and the Bb2$_{1}$m structure continues to be as the most stable configuration. 
Although, as in the case of PS, we can observe that for U $\geq$ 1.4 eV, it favors a semimetallic state in which
the Dirac-like bands (at M$_{x}$) are pushed to higher energies, besides the bands get closer to the Fermi level near the M$_{y}$ for U $\geq$ 1.8 eV.

\subsubsection{Effects of the functional and structural degrees of freedom}
\begin{figure}[ht]
\includegraphics[width=0.49\textwidth]{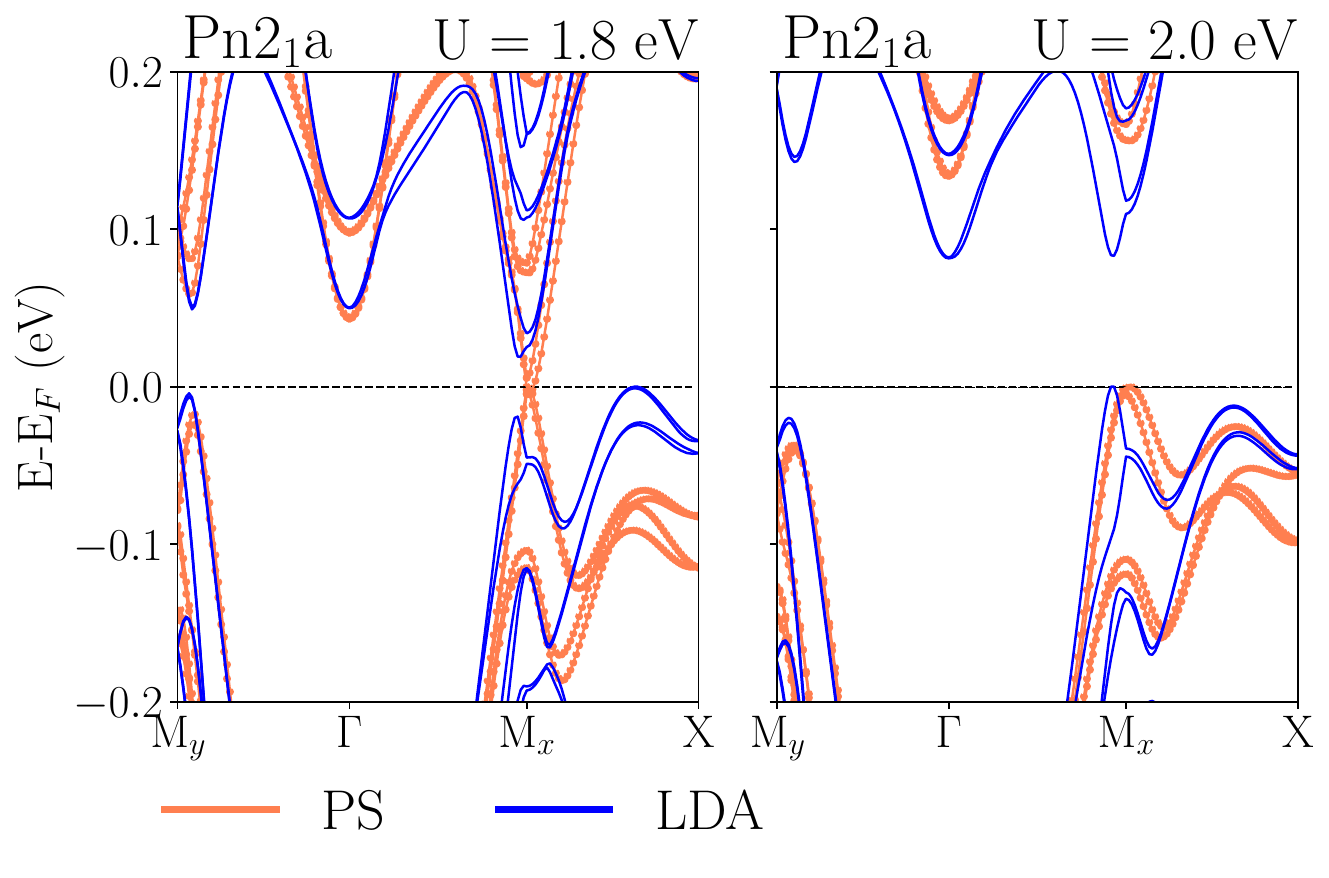}
\caption{Band structure for Pn2$_{1}$a structure performed with PS (color bands) and LDA (blue bands) approximations with U $=$ 1.8 eV (left panel) and U $=$ 2.0  eV (right panel).}
 \label{fig:PSvsLDA}
\end{figure}

%
To determine the mechanism promoting the electronic and structural changes under PS approximation, we perform an additional calculation to distinguish the effects of the functional choice and the structural phase in the electronic band structure. For this sake, first, we take the structure obtained with PS for U $>$ 1.4 eV (Pn2$_{1}$a phase), and with a frozen geometry, we calculate the electronic bands within the LDA approximation. In Fig. \ref{fig:PSvsLDA}, we compare the band structure for PS and LDA, keeping the geometry of the PS approximation (Pn2$_{1}$a). In the left panel of Fig. \ref{fig:PSvsLDA}, we show the band structure for U $=$ 1.8 eV, the system with Pn2$_{1}$a  (PS approximation) remains with a semimetal state. Instead, we can observe that the Dirac-like bands are suppressed under LDA approximation, emerging a band gap of around M$_{x}$-point. Furthermore, for U $=$ 2.0 eV, shown in the right panel of Fig. \ref{fig:PSvsLDA}, both approximations show a narrow bandgap insulator character with a gap at M$_{x}$ of 0.08 eV for LDA and 0.18 eV for PS approximation. Secondly, to reveal the effect of the structural degrees of freedom in the band structure for the Bb21m phase, we have calculated this without and with structural degrees of freedom along with PS and LDA approximation. In the first case, we find that the main role of Hubbard-U interaction is to control the band occupation around the Fermi level, favoring the gap increases around $\Gamma$ and the emergence of hole-like bands close to M$_{y}$ (for U $>$ 1.4) while changing the Hubbard-U from 0 to 2 eV (see SM. Fig. \ref{band-Bb21m}). In the second case,
the changes in the lattice vectors induce slight modifications of the bands around the Fermi level, affecting mainly the gap at $\Gamma$ (see SM Fig. S10-S11). In particular, the metallic character is still present for U = 1.4 eV within the PS approach; thus, the structural transition is not concurrent with the electronic transition from metal to semi-metal (gap at $\Gamma$ point).


\subsubsection{Comparison with Experimental Evidence}

In this section, we aim to get into contact with the experiments;
therefore, we focus on which approximation and parameters best describe the band structure and magnetic configuration (AFM-$b$) experimentally measured in CRO samples.
As discussed in section \ref{sec:adv}, understanding physical properties have been a significant challenge. There are inconsistencies between different experimental measurements and theoretical predictions. One of the main inconsistencies between observations and theory was that band structure calculation performed by DFT calculations predicted that CRO is a semi-metal with a high carrier concentration \cite{spindensityCa3}.
However, on the one hand, single-crystal measurements characterized CRO as a low carrier-concentration material on the verge of a metal-insulator. On the other hand, the ARPES measurements reveal that CRO displays a low-carrier metallicity but is characterized by a semimetallic state with a gapped fermi surface \cite{markovic2020electronically,horio2021electronic}.  A summary of contradictions around CRO can be found in ref. \citen{xing2018existence}.

A previous theoretical study (ref.\citen{danilo}) proposed a mechanism to model the band structure and Fermi surface of CRO at low temperatures based on DFT calculations (along with PS approximation). 
The suggested mechanism aim to explain the electronic changes between T$_{S}$ $<$ T $<$ 30 K, taking into account Hubbard-U in the range 0.5 $<$ U $<$ 1.6 eV, in which at U $=$ 1.4 eV  emerges a gap at $\Gamma$ point, accompanied of a spontaneous structural transition from Bb2$_{1}$m to Pn2$_{1}$a structure. In such scenario, the emergence of the semimetallic state could be related to the gap evidenced below 45 K in the optical conductivity spectra\cite{lee2007pseudogap}. Besides, the broken translational symmetry of the Pn2$_{1}$a structure could allow the charge (spin-) density wave transition agreeing with some experimental proves \cite{spindensityCa3}.

However, recently a series of independent experiments have successfully reconstructed the Fermi surface of CRO\cite{markovic2020electronically,horio2021electronic,xing2018existence} offering a novel approach to understanding the evolution of the electronic structure as the temperature decreases. 
The experimental evidence shows that CRO presents a semimetallic character at low temperatures\cite{markovic2020electronically,horio2021electronic}, in which the band structure displays a gap around $\Gamma$ point, Dirac-like bands at ($\pi$/a,$\pi$/b), and bands intersecting the Fermi level near the M$_{y}$. 
Each work\cite{markovic2020electronically,horio2021electronic} reveals a key aspect to consider to identify the mechanism underneath the CRO phase transition. On the one hand, it was shown that the reconstruction of the Fermi surface is incompatible with the translational symmetry-breaking density waves\cite{horio2021electronic}. On the other hand, a band gap magnitude of $\sim$ 10 to 15 meV around the $\Gamma$ point was estimated, comparable to the pseudogap observed in optical spectroscopy spectra \cite{markovic2020electronically}. These observations bring us several constraints to impose on our calculations.

After a systematic study and comparison with the most recent experiments, our results indicate that the LDA+U approximation is the one that most faithfully represents the complex behavior exhibited by the CRO.
The LDA+U approximation successfully identifies the magnetic ground state (AFM-$b$ configuration) and the occupation bands in agreement with the recent experimental reports\cite{horio2021electronic}. 

The subtlest region is to model the band spectrum along the high symmetry lines, specifically near the $\Gamma$ and the M$_{y}$; this region is quite sensitive to U value; however, these regions are key, as we explained before the experiment evidences a pseudogap (around $\Gamma$) and hole pocket (near the M$_{y}$)
and Dirac-like bands (at M$_{x}$)\cite{horio2021electronic}, 
given a semimetallic character to the system. In contrast to the PS approach, LDA can modulate the band occupation at $\Gamma$-point in the studied range of U-values. For instance, the LDA+U with U $=$ 0.5 eV evidence a band gap of 96 meV at $\Gamma$; this value is in the same order of ARPES measurements\cite{markovic2020electronically}. 
Besides, the bands are getting closer to M$_{y}$ for U $\geq$ 1.8 eV; however, for this range of U, the band dispersion around $\Gamma$ is pushed down to lower energies, increasing the size gap. Additionally, we have identified that the main band contribution around the Fermi level comes from d$_{xy}$, d$_{xz}$ and d$_{yz}$ orbitals in which d$_{xy}$ does not cross the Fermi level for U $>$ 0.5 eV (see Fig. \ref{S-bnd-partial}) in agreement with the experimental reports \cite{horio2021electronic}.
In the case of PS, several issues are found as U increases: Below U $<$ 1.4 eV band crosses the Fermi level at $\Gamma$ point and above U $\geq$ 1.4 eV due to a spontaneous transition toward the Pn2$_{1}$a structure, being this structural phase reported theoretically until now \cite{danilo} in the best of our knowledge.\\ 

In order to dig into the role of the Hubbard-U approximation and how the double counting, as well as the directional dependence, can affect the observed results, we have performed the same calculations using the Liechtenstein approximation \cite{liechtenstein1995density} (along PS). A detailed analysis is provided in section V of the SM; here, we summarize the main findings. We have observed that for U $<$ 1.4 eV Dudarev and Liechtenstein approximations have the same trend. However, at U $\geq$ 1.4 eV along Dudarev happens, the spontaneous phase transition and with Liechtenstein, the Bb2$_{1}$m phase keeps stable. With this last approximation, the transition happens for U $\geq$ 2.0 eV, emerging a narrow insulator phase.\\

\vspace{0.5cm}
\textbf{What is the underlying mechanism for the observed structural transition when considering the PS approximation with spin-orbit coupling (SOC) and Hubbard-U repulsion effects?}. 

The spontaneous transition appears independent of the Hubbard-U approximation. To inspect the role of SOC we have performed calculations considering Hubbard-U and structural degrees of freedom without SOC.
We find that without SOC, the spontaneous structural transition does not appear. Our calculations show that Bb2$_{1}$m is stable considering positions and volume degrees of freedom. However, the phase Pn2$_{1}$a is unstable, being difficult to have an accurate convergence. These results disagree with the studies reported in ref.\cite{danilo}. 

When spin-orbit coupling effects are considered, the direction of magnetization changes from point to point, inducing a complex magnetic landscape. The PBEsol approximation belonging to the family of generalized gradient functional approximations (GGA) is more susceptible to convergence problems and can induce symmetry changes to stabilize the system\cite{desmarais2019spin}.



\subsubsection{Linear response determination of U-term}

\begin{figure}
{\includegraphics[width=0.49\textwidth]{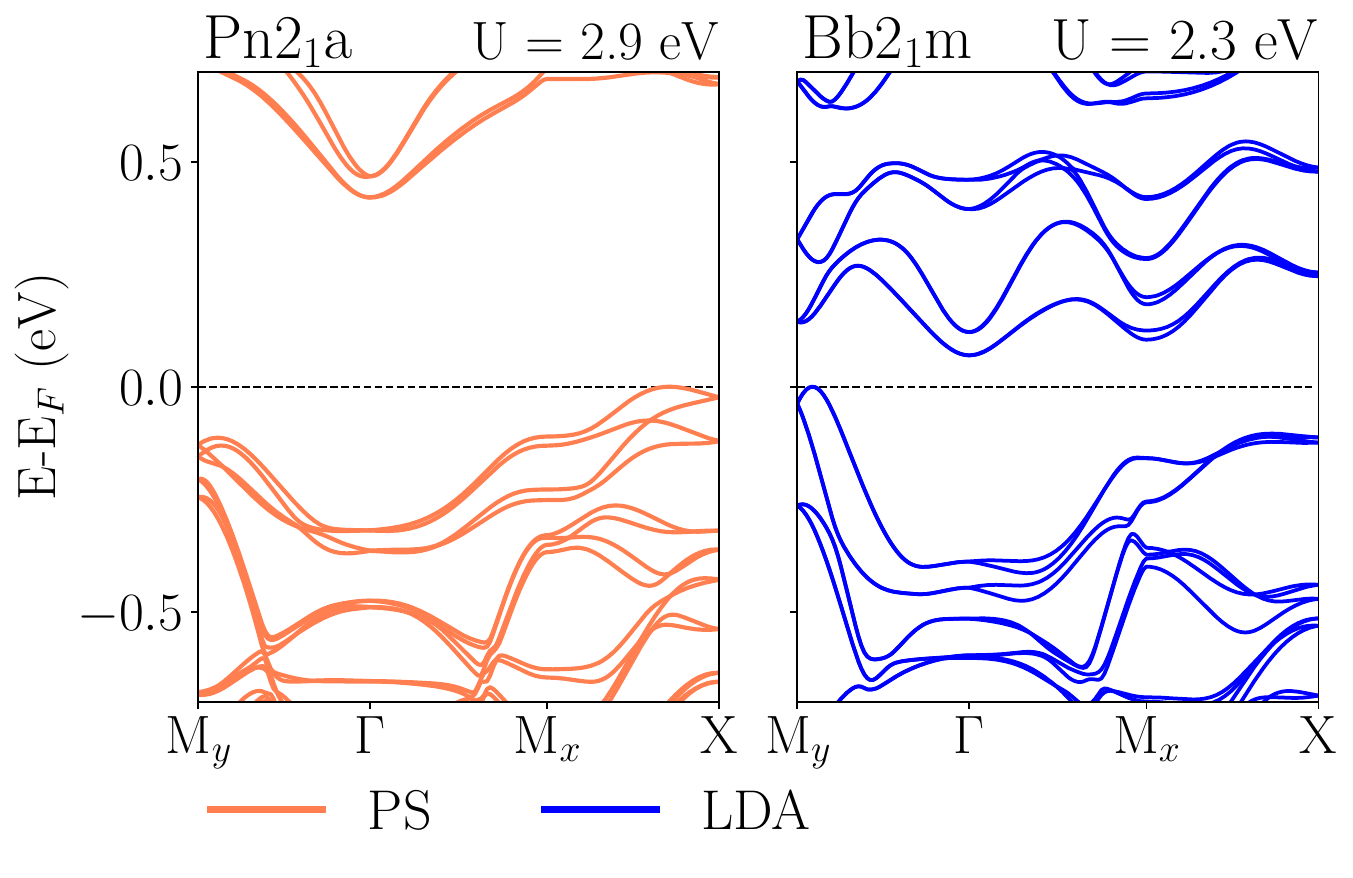}}
\caption{Band structure using U computed through the linear response method: PS with Pn2$_{1}$a structure for U$_{PS}$ $=$ 2.9 eV (left panel) and LDA with Bb2$_{1}$m phase using U$_{LDA}$ $=$ 2.3 eV (right panel).}
\label{bn-2.9}
\end{figure}
Going one step further, we employ a perturbative approach to determine a value for the U-Hubbard interaction. However, this method tends to overestimate the value of U. Note that the DFT+U approach also has limitations and should be considered a pragmatic way to partially eliminate the self-interaction errors inherent in DFT. In addition, the effective U-value depends on several factors that limit their transferability, such as atomic coordination and calculation parameters\cite{lim2016improved, yu2014communication} (pseudopotentials, basis-set).
We compute the U-Hubbard term through a constrained DFT-based method developed by Cococcioni and Gironcoli\cite{cococcioni2005linear} to compute the non-zero second derivative of energy concerning the local occupancy of the metal ion, in which $U$ is given by\cite{vaspwiki}:
\begin{equation}
    U\approx \left(\frac{\partial n^{\mathrm{scf}}_{I}}{\partial V_{I}} \right)^{-1}- \left(\frac{\partial n^{\mathrm{nscf}}_{I}}{\partial V_{I}}\right)^{-1},
\end{equation}
where $\partial{n_{I}} / \partial V_I$ is the occupation response at the site $I$, and V$_{I}$ is the orbital energy shift in the $I$ site. The first and second terms are the self-consistent (scf) and non-self consistent (nscf) solutions. We consider a perturbation $V$ in the $\pm$ 0.2 eV range. As a result of these calculations we found that $U_{\textrm{PS}} = 2.9$ eV and $U_{\textrm{LDA}} =2.3$ eV for PS and LDA respectively. We have also performed calculations in a lower range of perturbation ($\pm$0.1, with steps $\pm$ 0.02). Here, we have evidenced no large modification of the U value of U$_{LDA}$= 2.86 eV and U$_{ps}$= 2.73 eV.\\

We plot the electronic structure for the Hubbard-U values found by the linear response method. In Fig. \ref{bn-2.9} (left panel), the PS band structure ($U_{\textrm{PS}} = 2.9$ eV) and in Fig. \ref{bn-2.9} (right panel),  the LDA band structure ($U_{\textrm{LDA}} =2.3$ eV).
As a result of our calculation, we find that LDA retains the same Bb2$_{1}$m symmetry, and the PS approximation keeps the Pn2$_{1}$a phase. In the PS approximation, the larger U-term value ($U_{\textrm{PS}} = 2.9$ eV) produces a band gap of 0.80 eV at M$_{x}$  (see Fig. \ref{bn-2.9} left panel). Note that the band gap increases from the 0.153 eV calculated with U = 2 eV (outermost right panel of Fig. \ref{fig-band-u}). The large U influence in Pn2$_{1}$a structure induces t$_{2g}$ band distribution mainly at M$_{x}$, in which the band structure change from the hybridized bands d$_{xz}$-d$_{yz}$ for U $=$ 2.0 eV to be composed mainly by d$_{xz}$ orbitals (below Fermi-level) for U = 2.9 eV (see Fig. \ref{Fig-u20-ps}).

\begin{figure}
{\includegraphics[width=0.45\textwidth]{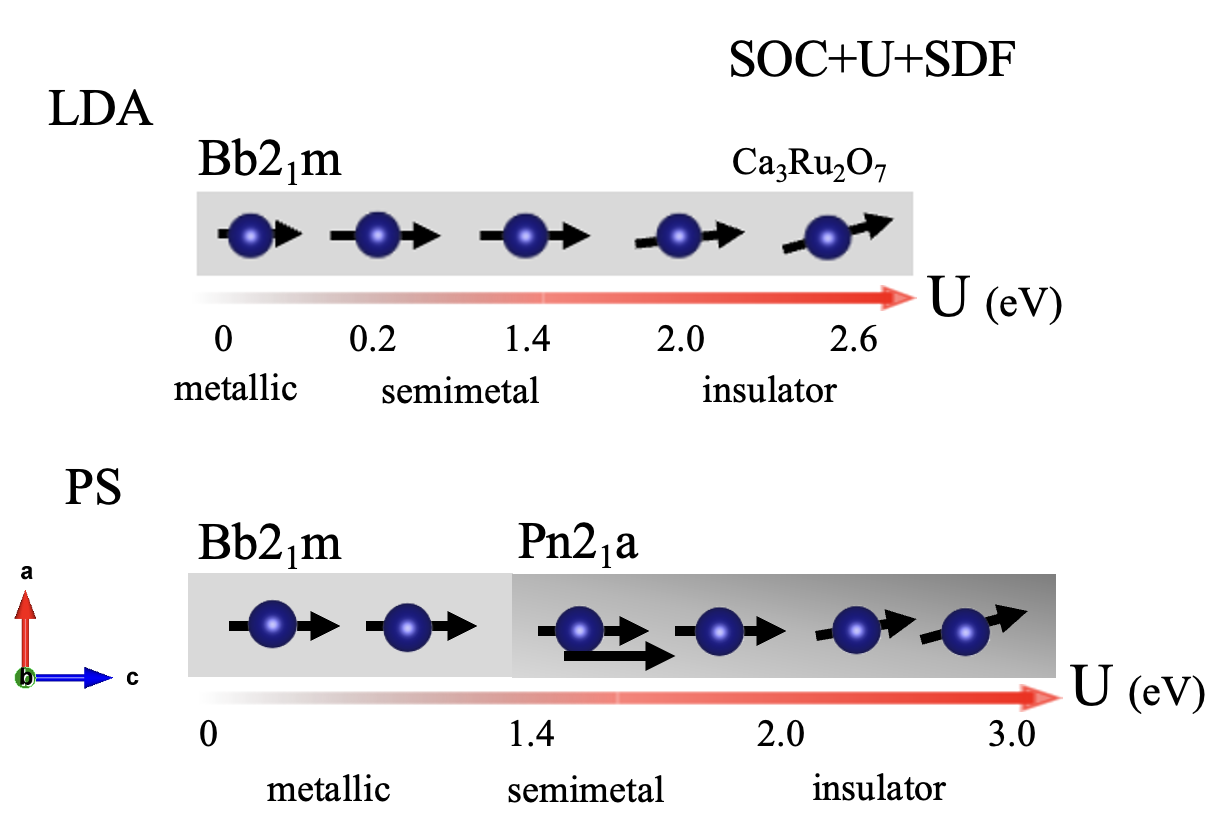}}
\caption{A schematic representation with the electronic and structural phase transitions as a function of Coulomb repulsion (U) in the presence of spin-orbit coupling (SOC) and structural degrees of freedom (SDF). The black arrows symbolize the spin direction of the Ru atoms (blue balls). As U increases, we can distinguish two main regions: among 0 $<$ U $\leq$ 2 eV in which the spins are along $b$ direction (AFM-$b$), and for U $>$ 2 eV, the spins turn aligned through (111) direction, in which the magnetic moment along $b$ direction is bigger than in $a$-$c$ direction (m$_{b}$ $>>$ m$_{a,c}$).}
\label{idea-phase}
\end{figure}

In the case of LDA approximation, the system presents a Bb2$_{1}$m symmetry; however, we can observe that to increase to U $=$ 2.3 eV, the Dirac-like bands are suppressed, emerging a gap of 0.10 eV at M$_{x}$ similar to the evidence along with PS approximation at U $=$ 2.0 eV (see Fig. \ref{fig-band-u}). The observed gap emerges due to the re-arrangement of t$_{2g}$ orbitals around M$_{x}$, as under the PS approximation case. Furthermore, in Bb2$_{1}$m structure the  large U induces a decrease of the d$_{yz}$ orbital population (see Fig. \ref{S-bnd-partial}) and for U $=$ 2.3 eV is favored the band occupation of d$_{xy}$-d$_{xz}$ around M$_{x}$ below the Fermi level (without hybridization, see Fig. \ref{S-bnd-partial} at U $=$ 2.3 eV).
Additionally, for larger values of the Hubbard U-term (U $>$ 2 eV), the magnetic state along the $b$ direction became unstable, and the magnetic moment direction goes toward the (111) direction in which m$_{b}$ $>$ m$_{a,c}$. Fig. \ref{idea-phase} shows a schematic representation with the electronic and structural phase transitions as U increases to 2.6 eV and 3.0 eV for LDA and PS approximations.

Because the structure of the ground state depends on the functional used (PS or LDA), a direct comparison between band structures is meaningless. To compare the band structure for a particular value of the Hubbard-U, we fix the atomic position and lattice vectors to the Bb2$_{1}$m structure (with the atomic position given by the experiment\cite{yoshida2005} at 8 K). Here, we noticed that regardless of the functional, the gap at M$_{x}$ appears for $U = 2.3$ eV; however, the system continues metallic (see Fig. \ref{PS-LDA-largeU}). For LDA approximation, the narrow bandgap insulator state appears for U above 2.3 eV (see Fig. \ref{PS-LDA-largeU} at U $=$ 2.6 eV). However, under PS approximation, the band structure favors the metallic character (see SM, Fig. \ref{PS-LDA-largeU}). The metallic behavior tendency could be related to a topological band arising as the U-term increases.

We should point out that the narrow bandgap insulator character originates in the strength of $U$-term independent of the functional choice and structural phase (in the presence of structural degrees of freedom). The Pn2$_{1}$a structure favors the insulator state for U $\geq$  2.0 eV and U $\geq$  1.8 eV for PS and LDA approximation, respectively (see Fig.\ref{fig:PSvsLDA}), whereas the Bb2$_{1}$m favors the insulator character for large U-term value, as we can observe for U $=$ 2.3 eV Fig.\ref{bn-2.9} and Fig. \ref{S-bnd-partial} under LDA approximation.

The large difference in the U-term computed through the linear approximation with the values chosen a priory could be related to the DFT self-interaction errors or $d$-ligand interactions between Ru and O atoms. To develop a U-Hubbard correction for the CRO system, it could be necessary to evaluate $d$-interactions for first principles, then construct a Hamiltonian that explicitly contains the $d$-orbitals and the $p$-orbitals of the system as interacting degrees of freedom; however, to solve this in practice could be a tremendous open challenge\cite{long2020evaluating, yu2014communication}.\\

To enclose our last finding, we suggest that the optimum values of CRO lie among 0.2 $\leq$ U $\leq$ 2.0 eV (under LDA approximation); in this range, the band structure agrees with the experimental report. However, for U $\geq$ 1.4 eV, the band dispersion is strongly pushed to lower around the $\Gamma$ point. The established range of U are in agreement with the U implemented to modulate the electronic structure in several ruthenium family such as: Ca$_{2}$RuO$_{4}$ (1.0 $\leq$ U $\leq$ 1.5)\cite{keen2021ab}, Sr$_{2}$RuO$_{4}$ (U $=$ 1.5 eV)\cite{iwasawa2010interplay}, CaRuO$_{3}$ (1.0 $<$ U $<$ 2.5 eV)\cite{zhong20104} in which both electronic and structural properties depend of the chosen U value.


\section{final remarks}
We have evidenced the strong interplay among magnetism, SOC,  U-Hubbard, and structural degrees of freedom on the electronic properties of CRO, considering two functional choices: LDA and PS approximations.

Depending on the level of theory, we find contradictory results. On the one hand, the LDA approximation fails to reproduce the correct magnetic ground state of CRO, predicting the AFM-$a$ solution as the most stable AFM configuration; however, the LDA+U approximation for U $\geq$ 0.2 eV results in the AFM-$b$ alignment as the ground state solution. On the other hand, PS and PS+U approximations successfully describe the AFM-$b$ magnetic configuration as the ground state.

Both LDA+U (0.2 $\leq$ U $\leq$ 2.0 ) and PS+U (0.5 $\leq$ U $\leq$ 1.6 eV) approximation successfully describe the electronic structure of CRO. However, regarding the structural properties, LDA predicts Bb2$_{1}$m phase as the ground state, and PS predicts a structural transition from Bb2$_{1}$m to Pb2$_{1}$a (U $\geq$ 1.4 eV), being this last one not evidenced by experiment so far. This new structural transition gives interesting explanations for describing several experimental findings, as explained in section-4. However, new experiments will be necessary to support the Pn2$_{1}$a structure evidenced using PS+U approximation.

We conclude that the exchange and correlation functional affects the band occupation mainly at $\Gamma$-point. On the one hand, the band crosses the Fermi level for PS approximation. On the other hand, a band gap appears for the LDA approximation. 
Additionally, the primary effect of the Coulomb repulsion (expressed as U-term) is to modify the band occupation around the Fermi level with the following main characteristics associated with the exchange-correlation approximation: 

\textbf{PS+U:} It is possible to distinguish three regions as the Hubbard-U term increases. The system exhibits a metallic character for U $<$ 1.4 eV, characterized by a band crossing the Fermi level  at $\Gamma$ and Dirac-like bands at M$_{y}$. For U $\geq$ 1.4 eV, the Bb2$_{1}$m phase became unstable, and the system relaxes spontaneously into Pn2$_{1}$a structure, exhibiting a semimetallic state (gap around $\Gamma$ point) (independent of the structural degrees of freedom considered). Finally, for U $\geq$ 2.0 eV, a semimetal to a narrow bandgap insulator phase emerges by opening a gap of 0.16 eV at M$_{x}$ (U $=$ 2.0 eV). 

The observed structural transition is independent of the U approach; along Liechtenstein approach, we find that the transition emerges for U $>$ 2.0 eV together with an electronic transition from semi-metal to narrow-insulator state (in agreement with PS approximation for U $\geq$ 2.0 eV).\\

\textbf{LDA+U:} The Bb2$_{1}$m phase is the most stable configuration, and the band structure presents a semimetallic character. For 0.2 $\leq$ U $\leq$ 2.0 eV, we have evidenced a gap around $\Gamma$ (gap size of 96 meV at $\Gamma$ for U $=$ 0.5 eV) and Dirac-like bands at M$_{x}$ in good agreement with the band structure reported by the experiments. However, for above U $\geq$  1.4 eV, the band around $\Gamma$  is pushed down to lower energies increasing the gap size (as also happens with PS). 

We remark that a narrow bandgap insulator state emerges independent of the functional choice. In the case of PS+U happening for U $\geq$ 2.0 eV (Dudarev scheme) and U $>$ 2.0 eV (Liechtenstein approach). For LDA+U occurring from U $>$ 2.0 eV (Dudarev scheme). Additionally, for these large U values, the magnetic moment along the $b$ direction became unstable, and $m$ goes from [010] to [111] direction in which m$_{b}$ $>$ m$_{a,c}$. 

From these findings, we can give keys to model the electronic structure of CRO system, establishing the effect of Coulomb repulsion and structural degrees of freedom for two different functional natures. Here we have evidenced that CRO is very sensitive to these parameters, in which, as a function of U, it is possible to get a metal, semimetal, and narrow bandgap insulator state independent of the structural degrees of freedom. 
Depending on the functional choice Bb2$_{1}$m phase can be stable (LDA) or unstable (PS) in the presence of Coulomb repulsion, SOC and structural degrees of freedom.\\

The Hubbard repulsion term is usually considered a free parameter, adjusted to improve agreement between calculated and measured properties. This is one of the simplest approaches used to improve the description of the ground state of the many correlated systems, based on the correction of E$_{LDA/GGA}$ energies, leading to the called ``double-counting term (E$_{dc}$)" that models the contribution of correlated electrons to the DFT energy as a mean-field. However, the dc term is not uniquely defined, being this an open issue of LDA+U that has been widely discussed; see ref. \cite{ostlin2016electronic} for further discussion. Some techniques, such as the linear response approach, can estimate a value for the effective U-term; these techniques, however, require further validation. Methods beyond DFT, such as Local Density approximation + Dynamical Mean-Field approximation (LDA+DMFT), have been successfully used to describe the electronic, structural properties and phase transitions in correlated materials with 4d electrons \cite{ostlin2016electronic}, including Ruddlesden-Popper compounds \cite{ca-strain1,ca-strain2}. Studies in this direction could provide a better understanding of CRO, making it possible to compare different levels of theory on this compound. In this sense, there is still much to explore in CRO systems. In this direction, our study aims to give a first step forward, doing a systematic study of the role of the exchange and correlation in the presence of different degrees of freedom and considering different approximations for the Hubbard-U parameter.

While preparing this manuscript, we became aware of a recent experimental report by Hao-Yu et al. (ref.~\citen{niu2022magnetoresistance}) studying the magnetoresistance and magnetostriction on CRO. This study proposes the use of pulsed magnetic fields to control the magnetic states of CRO. Additionally, in agreement with our results, they reported the stability of magnetic phases (AFM-b, AFM-a) using PBEsol+U with $U = 1.2$ eV. Furthermore, a recent experimental report has studied the CRO under lattice deformation (ref.~\citen{dashwood2022strain}), evidencing that applied strain can be used to tune the spin orientation. These studies confirm the strong correlation between lattice and electronic degrees of freedom needed to reproduce the magnetic transitions of CRO.

\section{acknowledgment}
The authors would like to thank Igor Markovic and Jose Mej\'ia for the fruitful discussions.
We acknowledge the Max Planck Computing Data Facility for bringing the technical support and Dr. Claudia Felser for providing the VASP license. 
AL thanks to Chilean FONDECYT POSTDOCTORAL grant N. 3220505 and Dr. Danilo Puggioni for aiding with the calculation setup and discussions in the early stage of this project.
JWG acknowledges financial support from FONDECYT: Iniciaci\'on en Investigaci\'on 2019 grant N. 11190934 (Chile) and USM complementary research grant.


\bibliographystyle{unsrt}
\bibliography{bibliography.bib}

\pagebreak
\widetext

\newpage
\begin{center}

\textbf{\large Supplemental Material: Ca$_{3}$Ru$_{2}$O$_{7}$: Interplay among degrees of freedom and the role of the exchange-correlation}\\

Andrea León,$^{1,2,3}$ J. W. Gonz\'alez,$^4$, H. Rosner$^2$

$^1$Facultad de Física, Pontificia Universidad Católica de Chile, Santiago, Chile.\\
$^2$ Max Planck Institute for Chemical Physics of  Solids, Dresden, Germany.\\
$^3$ Faculty for Chemistry, TU Dresden, Bergstrasse 66c, 01069 Dresden, Germany.\\
$^4$ Departamento de F\'{i}sica, Universidad  T\'{e}cnica Federico Santa Mar\'{i}a, Casilla Postal  110V, Valpara\'{i}so, Chile.\\

\end{center}

\setcounter{figure}{0} 
\setcounter{section}{0} 
\setcounter{equation}{0}
\setcounter{table}{0}
\setcounter{page}{1}
\renewcommand{\thepage}{S\arabic{page}} 
\renewcommand{\thesection}{S\Roman{section}}   
\renewcommand{\thetable}{S\arabic{table}}  
\renewcommand{\thefigure}{S\arabic{figure}} 
\renewcommand{\theequation}{S\arabic{equation}} 

\vspace{0.5cm}
The next sections, I and II, show the structural and electronic properties of CRO considering, in all the cases, spin-orbit interaction and different Hubbard-U parameters, magnetic states, and with and without structural degrees of freedom. Section III compares the main electronic and structural properties of CRO, considering two approximations to include U-Hubbard repulsion: Dudarev and Liechtenstein approach. 

\section{Structural properties}

\subsection{Volume degrees of freedom.}
    Table \ref{S1-vol-is3} shows the lattice parameter in \AA, Ru-O length in \AA, octahedra volume and equilibrium volume in \AA$^{3}$ for PBEsol (PS), LDA approximation, and the values given by the experiment. Here, we can observe that LDA and PS approximation underestimate the volume by 4\%  and 1.44\% , respectively. Table \ref{S2-energy-is3} shows the energy difference between AFM-$a$ and AFM-$b$ magnetic configuration ($\Delta$E= E$_{AFM-a}$-E$_{AFM-b}$) per Ru atom and the magnetic moment per Ru atom ($m$) for each magnetic phase (calculation considering U $=$ 0 eV)
.

\begin{table}[H]
    \centering
    \begin{tabular}{c|c|c|c|c|c|c}
    \hline
& a (\AA) & b (\AA) & c (\AA) & Ru-O  (\AA) & V(RuO$_{6}$)  (\AA$^{3}$) & V$_{0}$ (\AA$^{3}$)\\
\hline
LDA    & 5.292 &  5.458  & 19.427   & 1.965& 10.09 & 556.72    \\
PS     & 5.328 & 5.533 & 19.390 & 1.986 & 10.44 & 571.63  \\
Exp.\cite{yoshida2005}   & 5.367 & 5.535 & 19.521 & 1.992 & 10.53  & 580.04    \\
\hline
\end{tabular}
    \caption{Equilibrium lattice parameter computed with LDA and PBEsol (PS) approximation. The experimental values are given by \cite{yoshida2005}.}
\label{S1-vol-is3}
\end{table}

\begin{table}[H]
 \centering
    \begin{tabular}{c|c|c}
\hline
Approximation & $\Delta$E (meV/Ru) &  $m$ ($\mu_{B}$)   \\
\hline
LDA    & -3.644    & 0.28  \\
PS    & 0.44625  & 1.34  \\
Exp\cite{yoshida2005}    & AFM-$b$    & 1.8  \\
       \hline
\end{tabular}
\caption{$\Delta$E is the energy difference between AFM-$a$ and AFM-$b$ magnetic configuration ($\Delta$E= E$_{AFM-a}$-E$_{AFM-b}$) per Ru atom and $m$ the magnetic moment per Ru atom.}
\label{S2-energy-is3}
\end{table}

\subsection{Hubbar-U and structural degrees of freedom.}

Table \ref{is2-leng-U} shows the equilibrium volume and magnetic moment for several U values for LDA and PS approximation.

\begin{figure}[ht]

\subfigure[]{\includegraphics[width=0.45\textwidth]{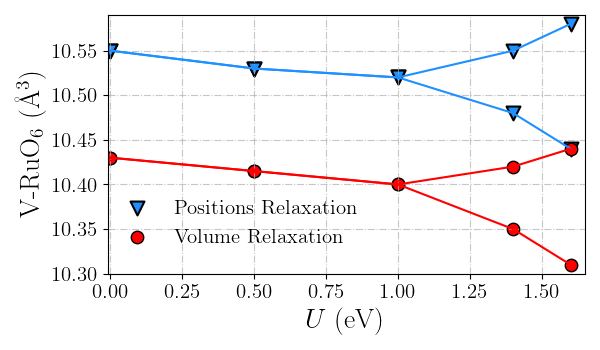}}
\subfigure[]{\includegraphics[width=0.45\textwidth]{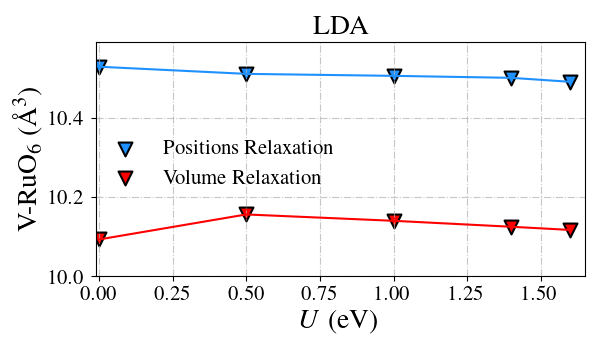}}
\caption{Volume octahedra (V-RuO$_{6}$). Calculation considering positions and volume degrees of freedom.}
\label{relaxed-nrelate}
\end{figure}

Fig.\ref{relaxed-nrelate} shows the RuO$_{6}$ volume octahedra as U increases under positions and volume degrees of freedom (see equilibrium volume in the Table \ref{is2-leng-U}). 

\begin{table}[ht]
 \centering
 \caption{PS and LDA approximation: Pressure, volume by cell, magnetic moment ($m$), for several U values (in eV) under volume degrees of freedom}
    \begin{tabular}{c|c|c|c|c|c|c}
\hline
 & U (eV) & 0 & 0.5 & 1.0 & 1.4 &  1.6  \\
\hline
 PS      & Volume Cell (\AA$^{3}$) & 571.62  & 571.23 & 570.70  & 570.26  & 570.16     \\
        & m ($\mu_{B}$) & 1.35 & 1.38 & 1.41 & 1.40/1.44 &  1.38/1.48   \\
\hline
LDA     & Volume Cell (\AA$^{3}$) & 550.72  & 554.34 & 554.07 &  553.63  & 553.36     \\
        & $m$ ($\mu_{B}$) & 0.28 & 1.27 & 1.35 & 1.37 & 1.38   \\
\hline
\end{tabular}
    \label{is2-leng-U}
\end{table}

\section{Electronic properties}

\subsection{FM phase: positions degrees of freedom.}

\begin{figure}[H]
\centering
{\includegraphics[width=0.35\textwidth]{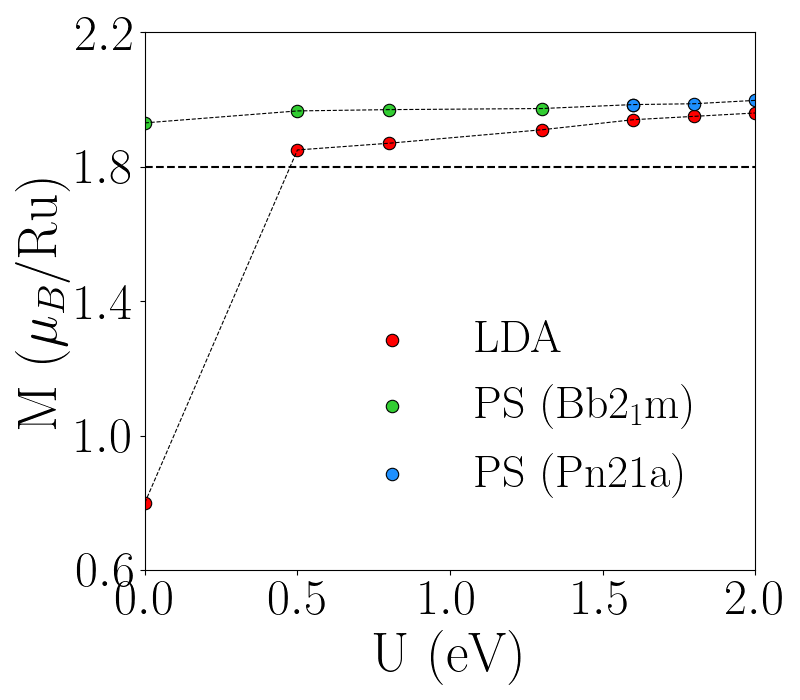}}
\caption{Magnetic moment total by Ru atom average with the magnetic moment projected along b direccion (FM-b), for LDA and PS approximation. Black dashed lines shows the experimental magnetic moment value \cite{yoshida2005}.}
\label{m-fm}
\end{figure}

The total magnetic moment (M) by Ru atom (defined as M = total magnetic moment of the cell/Ru) as U increases is given by the Fig.\ref{m-fm}.

\subsection{Band structure: positions degrees of freedom}

\begin{figure}[ht]
\centering
{\includegraphics[width=0.8\textwidth]{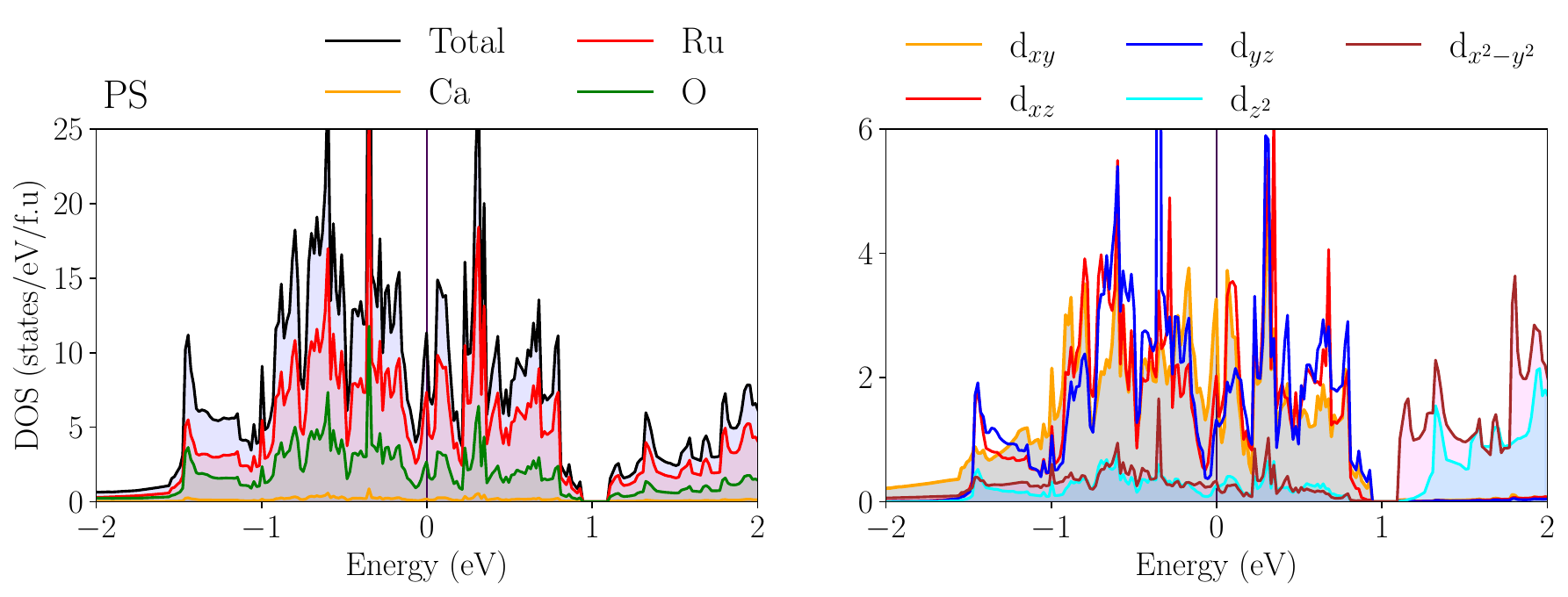}}
{\includegraphics[width=0.8\textwidth]{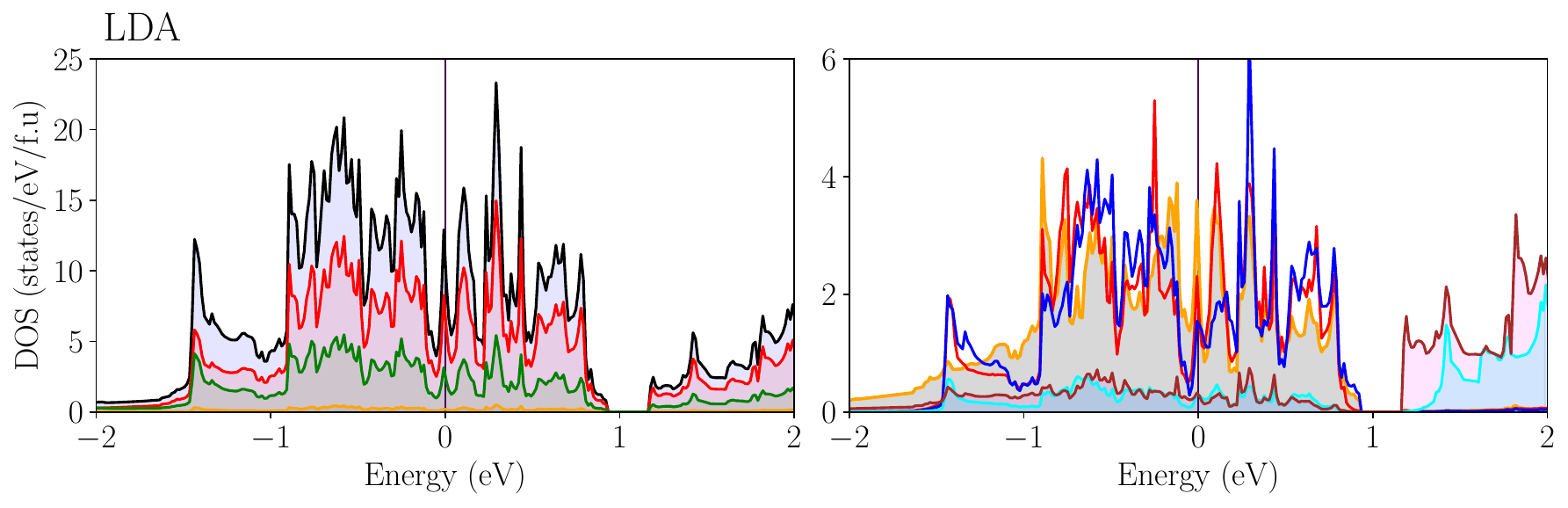}}
 \caption{Density of states (DOS) for AFM-$b$ configuration along PS and LDA approximation at U $=$ 0 and U $=$ 0.5 eV, respectively. The Fermi level is settled at E $=$ 0.}
\label{S3-lda-ps-DOS}
\end{figure}

Fig. \ref{S3-lda-ps-DOS} shows the density of states (DOS) along LDA and PS calculation at U $=$ 0 for PS and U $=$ 0.5 for LDA at the experimental volume.\\

\begin{figure}[ht!]
\subfigure[]{\includegraphics[width=0.4\textwidth]{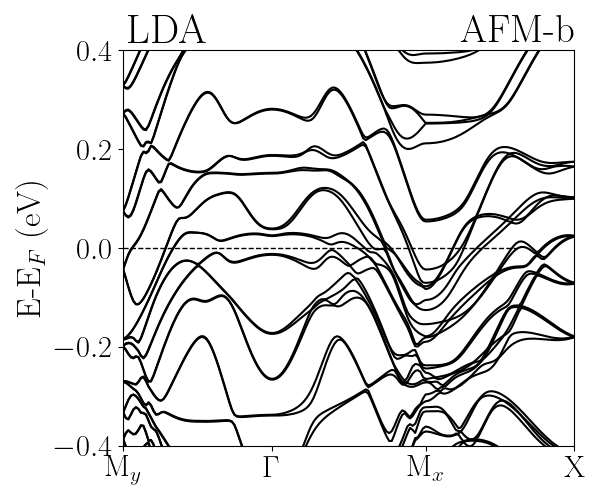}}\subfigure[]{\includegraphics[width=0.4\textwidth]{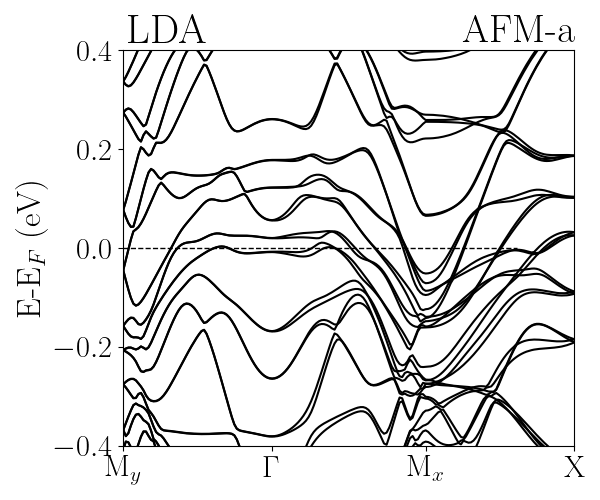}}
 \caption{(a)-(b) Band structure for LDA approximation with AFM-$b$ and AFM-$a$ configuration.}
\label{S4-lda-bands-A-B}
\end{figure}

Fig. \ref{S4-lda-bands-A-B} (a) and (b) show the band structure for LDA calculation with AFM-$b$ and AFM-$a$ configuration considering U = 0 eV at the experimental volume.

\begin{figure}[H]
{\includegraphics[width=0.5\textwidth]{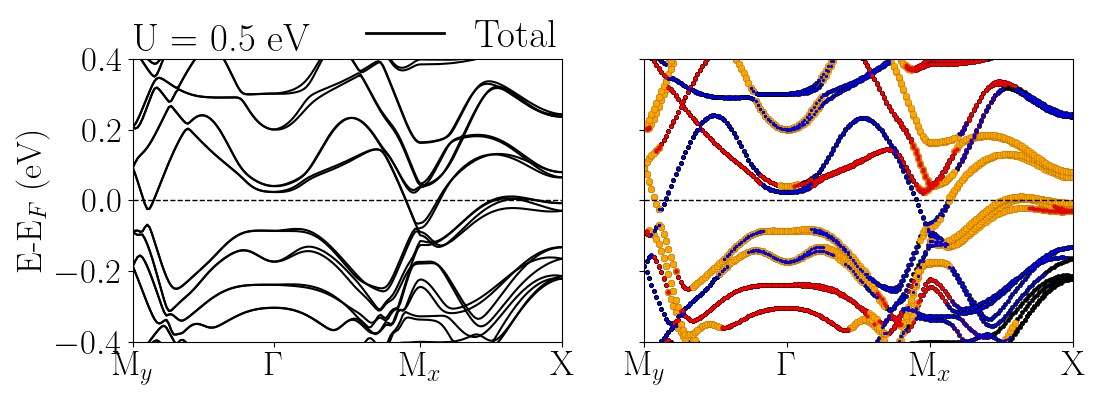}}{\includegraphics[width=0.5\textwidth]{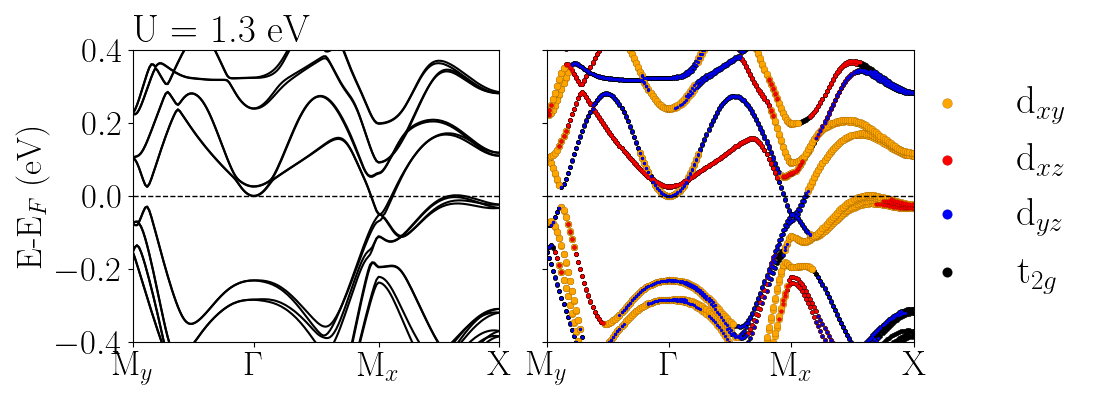}}
{\includegraphics[width=0.5\textwidth]{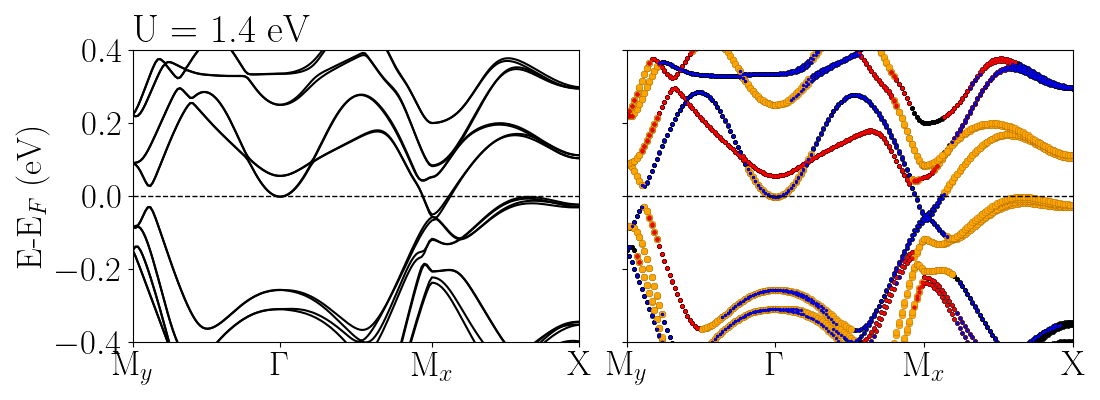}}{\includegraphics[width=0.5\textwidth]{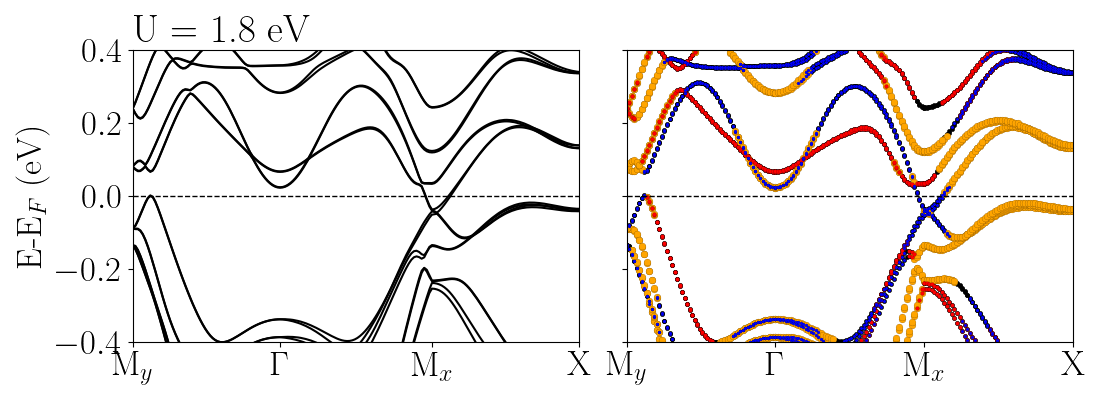}}
{\includegraphics[width=0.5\textwidth]{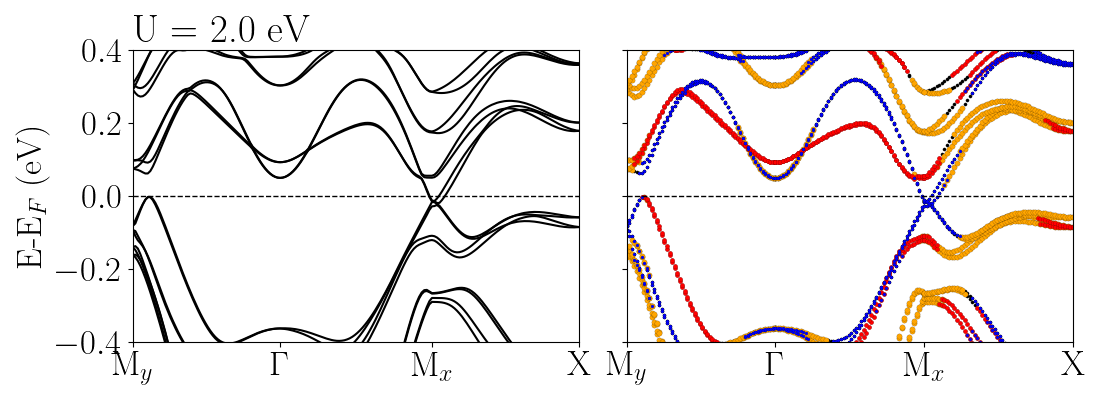}}{\includegraphics[width=0.5\textwidth]{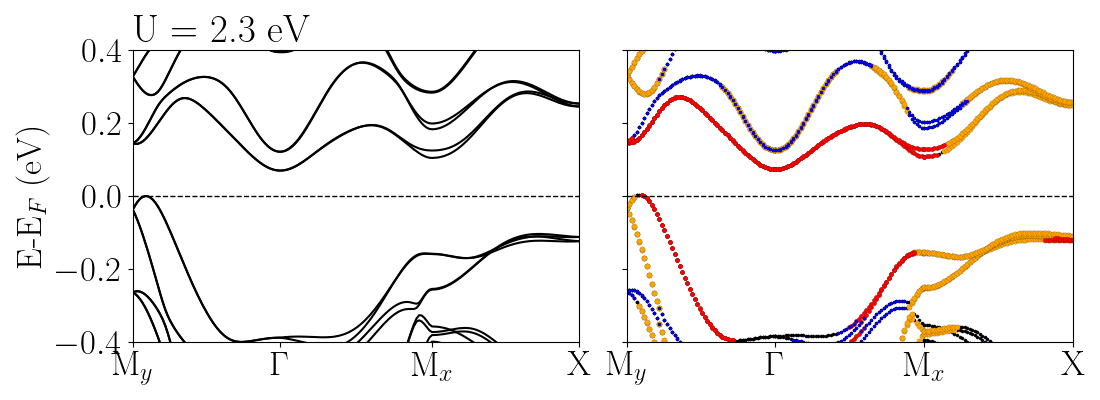}}
 \caption{Total and $d$-partial band and structure for AFM-$b$ configuration along LDA approximation for U $=$ 0.5, 1.3, 1.4, 1.8, 2.0 and 2.3 eV. }
\label{S-bnd-partial}
\end{figure}

\begin{figure}[ht]
{\includegraphics[width=0.5\textwidth]{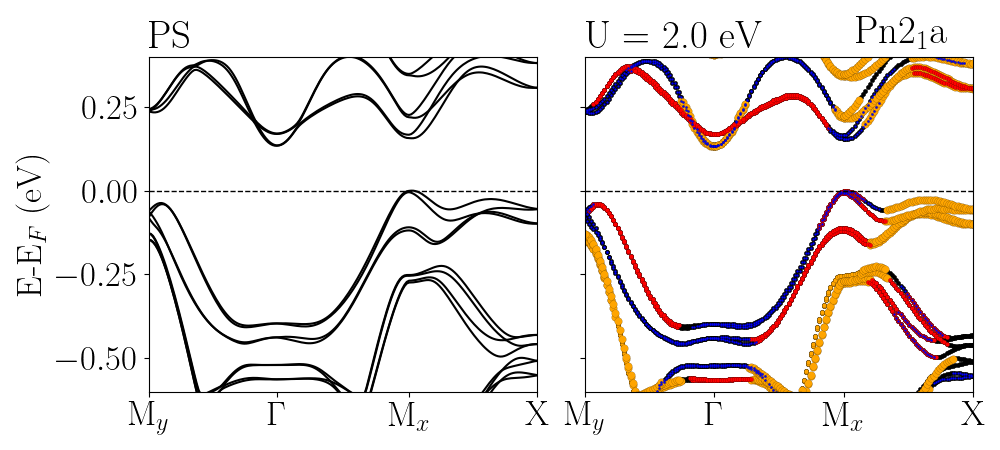}}{\includegraphics[width=0.5\textwidth]{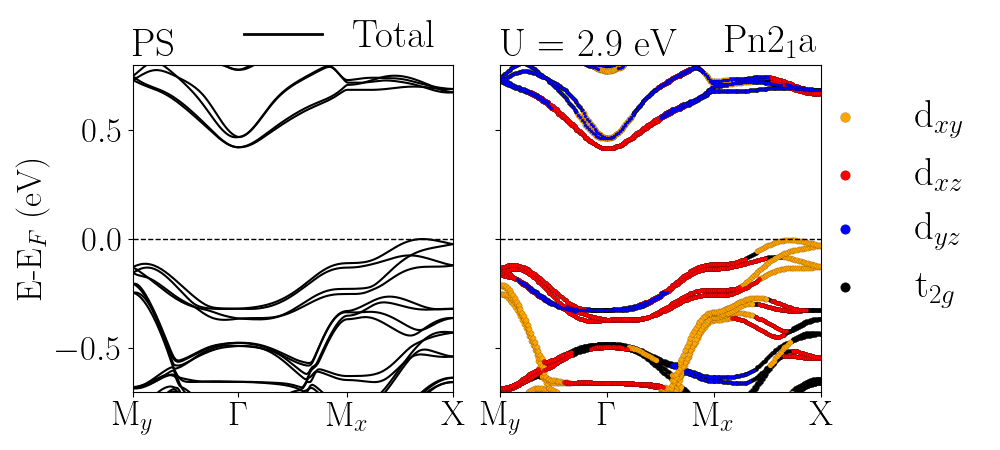}}
 \caption{Total and partial band structure for PS approximation for U $=$ 2.0 and 2.9 eV.}
\label{Fig-u20-ps}
\end{figure}

In the following the displayed $d$-partial band structure have a percentage of occupation between 50-75\% per each orbital; these correspond to
d$_{xy}$, d$_{xz}$ and d$_{yz}$, the d$_{x2-y2}$ and d$_{z2}$ represent less than 25\% of occupation then they are not shown. The back point displays the hybridized t$_{2g}$ orbitals representing between 0-25\% of the occupation per band (Fig. \ref{S-bnd-partial}, \ref{Fig-u20-ps}, \ref{PS-LDA-largeU}).\\

Fig. \ref{S-bnd-partial} shows the electronic band structure performed in the phase Bb2$_{1}$m for several U values.
Along with LDA approximation. Here we can observe that the $d$-electronic band distribution changes slightly 
to increase the U-term. Along M$_{y}$-$\Gamma$ bands are pushed to high energies inducing a band distribution, 
helping the occupation of the d$_{xz}$ orbitals, which slightly hybridize with the d$_{xy}$ orbitals 
just below the Fermi level. Around $\Gamma$ point, we can observe that the d$_{xy}$ contribution decreases, 
and these states become more hybridized with d$_{yz}$ orbitals below the Fermi level. Additionally, these 
bands are strongly pushed toward lower energies. At M$_{x}$, the Dirac-like bands move towards the Fermi 
level, favoring the occupation of d$_{yz}$ orbitals.
Finally, from M$_{x}$-X, we can observe that these states are slightly pushed below the Fermi level. 
These bands are rather modified because they are mainly composed of d$_{xy}$-orbitals being close to integer 
filling and then exhibiting a lower grade of hybridization. Furthermore, we have included the band structure at
U $=$ 2.3 eV  (U computed using the perturbative approach \cite{cococcioni2005linear}). For this U value, 
we can observe that a band gap of 0.25 eV emerges at M$_{x}$ arising the transition from metal to a narrow bandgap insulator. Together with the transition, we can observe a $d$-orbital re-ordering mainly at M$_{x}$ in which the 
dominant bands are mainly d$_{xz}$-d$_{xy}$. 

Fig. \ref{Fig-u20-ps} shows the total bands and partial $d$-band structure along with PS approximation for the U value in which happens the electronic transition from semimetal to a narrow bandgap insulator (U $=$ 2.0 eV) with Pn2$_{1}$a structure with a gap of 0.18 eV at M$_{x}$. Here, we can observe that around M$_{x}$ appears a grade of hybridization between d$_{xz}$ and d$_{yz}$ orbitals. To increase the U to 2.9 eV (U computed using the perturbative approach), we can observe that the gap increases to 0.80 eV at M$_{x}$. The bands around this point are mainly composed of d$_{xz}$ orbitals.\\

\begin{figure}[ht!]
\includegraphics[width=1.0\textwidth]{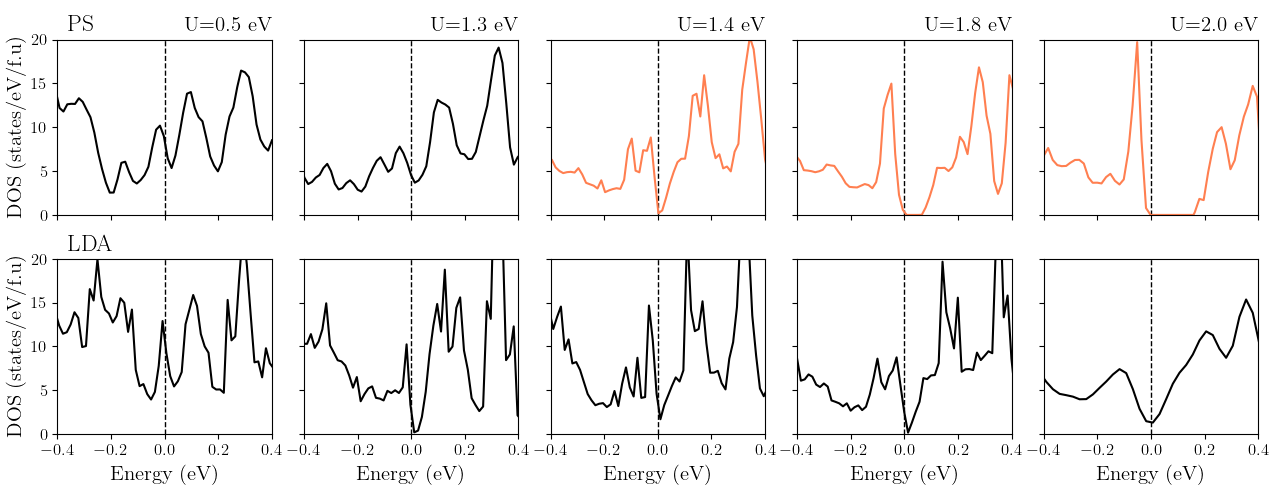}
\includegraphics[width=0.25\textwidth]{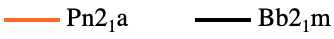}
 \caption{Total density of states (DOS) along PS and LDA approximation around the Fermi-Level (E $=$ 0). The black band corresponds to the Bb2$_{1}$m structure, and the color band to the Pn2$_{1}$a phase. Calculation considering SOC+U+Lattice degrees of freedom.}
\label{S-dos-u}
\end{figure}

Fig. \ref{S-dos-u} shows the total density of state (DOS) in a window of [-0.4,0.4] eV. For PS approximation, we can observe that for U= 0.5 and 1.3 eV, the system exhibits a metallic state with Bb2$_{1}$m structure. When  U increases to 1.4 eV-1.8 eV, the DOS at the Fermi level decreases, being close to 0, giving a nearly semimetallic character along with a Pn2$_{1}$a phase. For U= 2.0 eV is observed a gap opening at the Fermi level of 0.16 eV.
In the case of LDA approximation, as U increases, we can see a diminution of the states around the Fermi level; however, the system continues in a metallic phase with Bb2$_{1}$m structure.

\subsection{Band structure: without structural degrees of freedom}

\begin{figure}[H]
\includegraphics[width=1.0\textwidth]{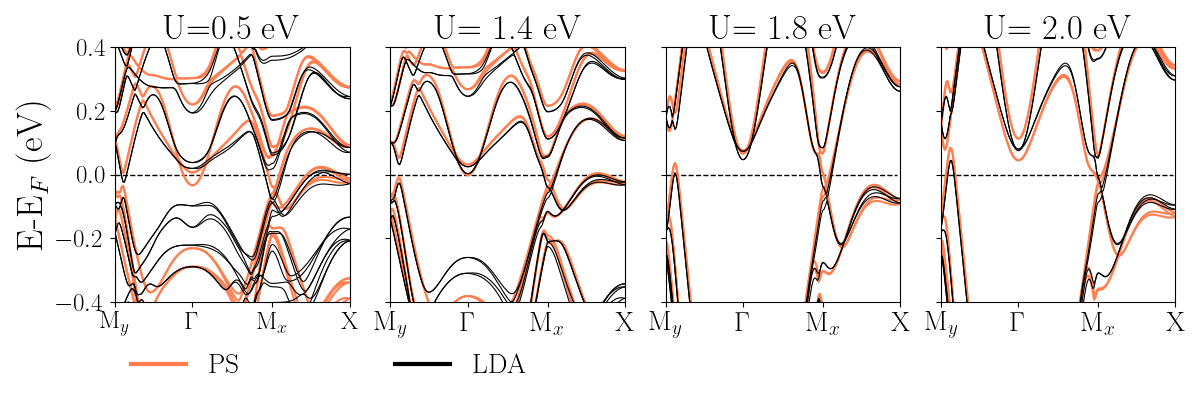}
\caption{Band structure considering Bb2$_{1}$m structure for U $=$ 0.5, 1.4, 1.8 and 2.0 eV. Black and orange lines correspond to LDA and PS approximation. Calculation without considering structural degrees of freedom.}
\label{band-Bb21m}
\end{figure}

Fig. \ref{band-Bb21m} shows the electronic band structure performed in the Bb2$_{1}$m structure (with the position reported by the experiment at 8 K\cite{yoshida2005}) for several U values without considering the lattice degrees of freedom. Here we can observe that the role of U is to control the band occupation mainly around the $\Gamma$ point (above and below the Fermi level). Furthermore, we evidence that without degrees of freedom, the band opening at M$_{x}$ at U $=$ 2.0 eV does not emerge for PS approximation (see Fig. \ref{fig-band-u} main text).

\begin{figure}[ht!]
\includegraphics[width=0.5\textwidth]{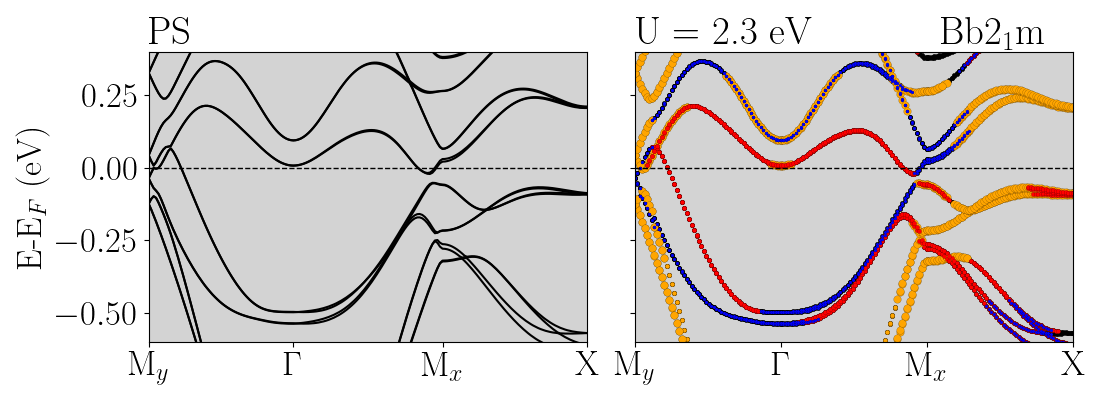}\includegraphics[width=0.5\textwidth]{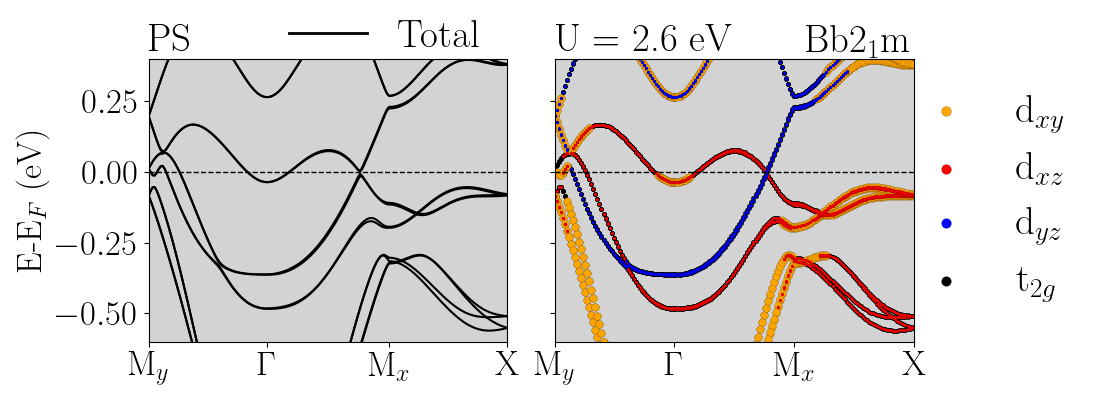}
\includegraphics[width=0.5\textwidth]{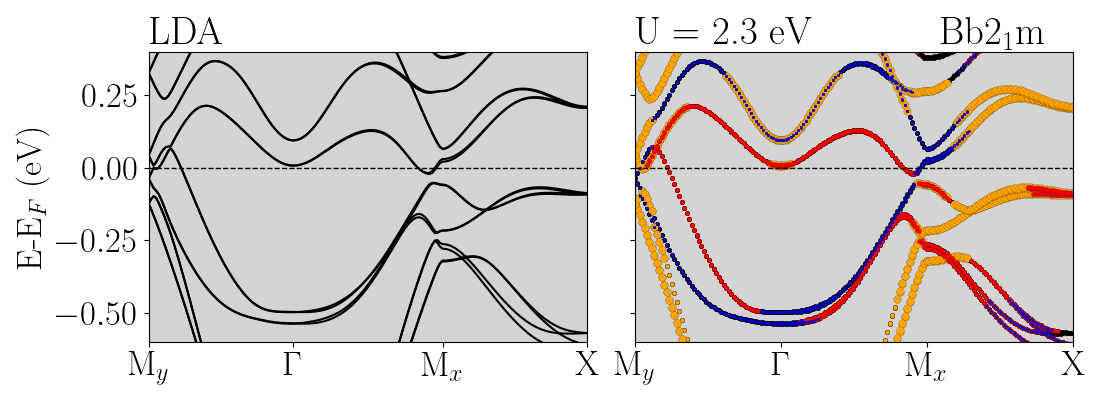}\includegraphics[width=0.5\textwidth]{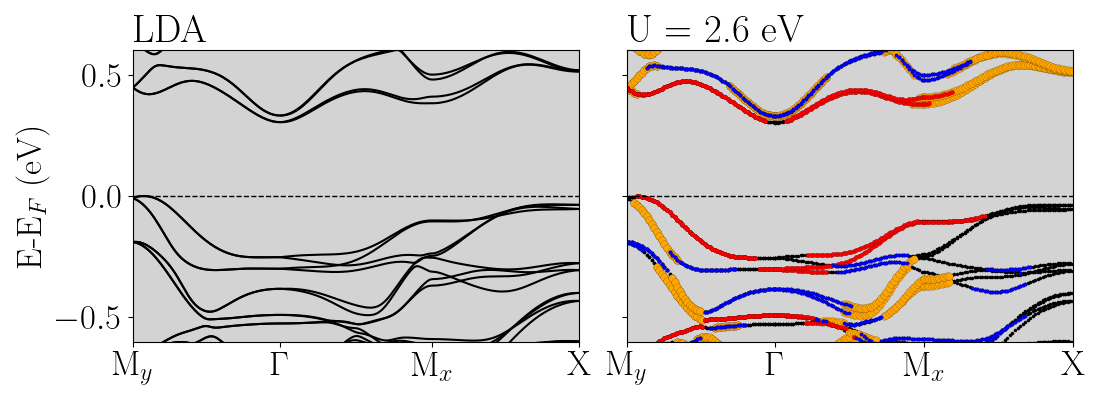}
 \caption{Band structure along PS and LDA approximation with Bb2$_{1}$m structure for U $=$ 2.3 eV and 2.6 eV (without structural degrees of freedom).}
\label{PS-LDA-largeU}
\end{figure}

Fig. \ref{PS-LDA-largeU} shows the total band and partial $d$-band along with PS and LDA approximation, with U $=$ 2.3 eV, considering Bb2$_{1}$m structure without considering structural degrees of freedom. Here we can observe that for both PS and LDA approximation emerges, a gap at M$_{x}$; however, the system continues in a metallic state. To increase U to 2.6 eV, we can observe that the system became a narrow bandgap insulator in the case of LDA. Instead, along PS, the system goes toward a metallic state.

\subsection{Band structure: structural degrees of freedom.}

\begin{figure}[H]
\centering
{\includegraphics[width=0.4\textwidth]{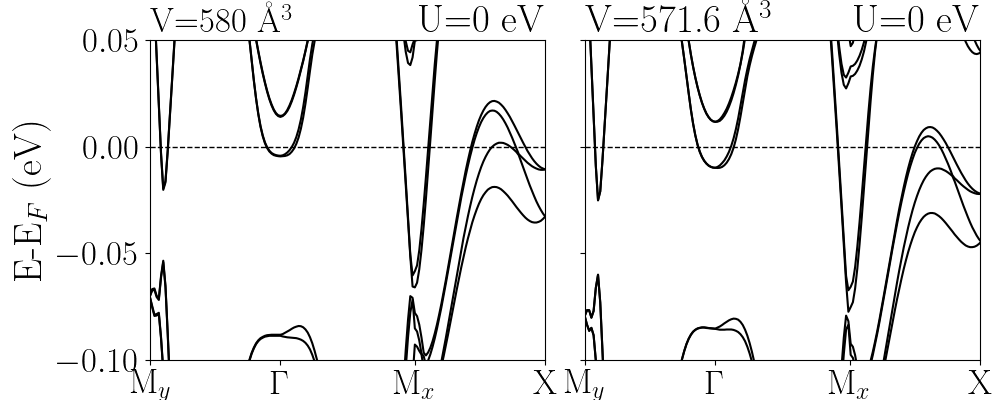}}{\includegraphics[width=0.4\textwidth]{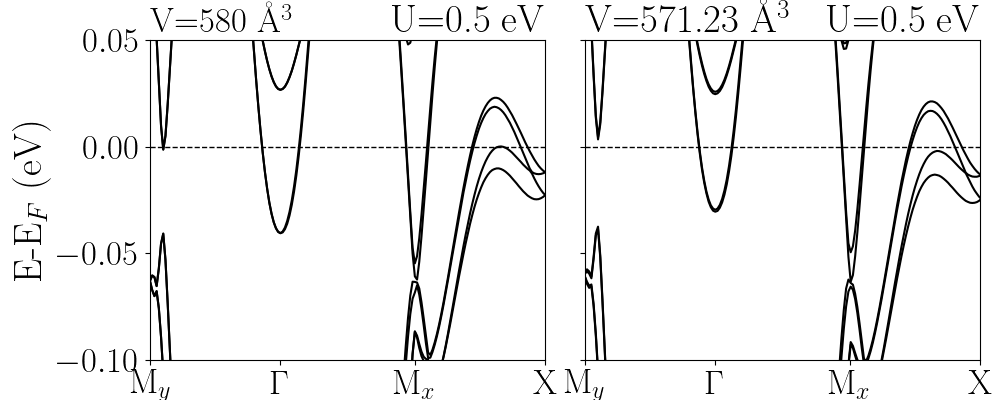}}\\
\vspace{0.2cm}
{\includegraphics[width=0.4\textwidth]{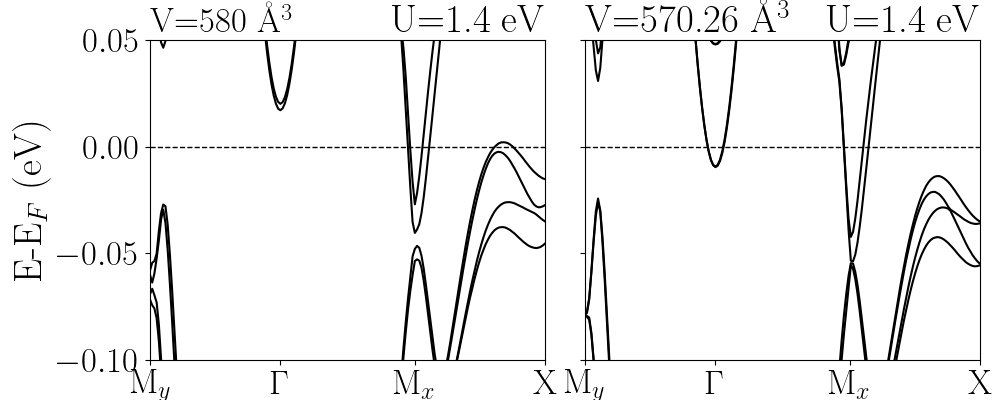}}
{\includegraphics[width=0.4\textwidth]{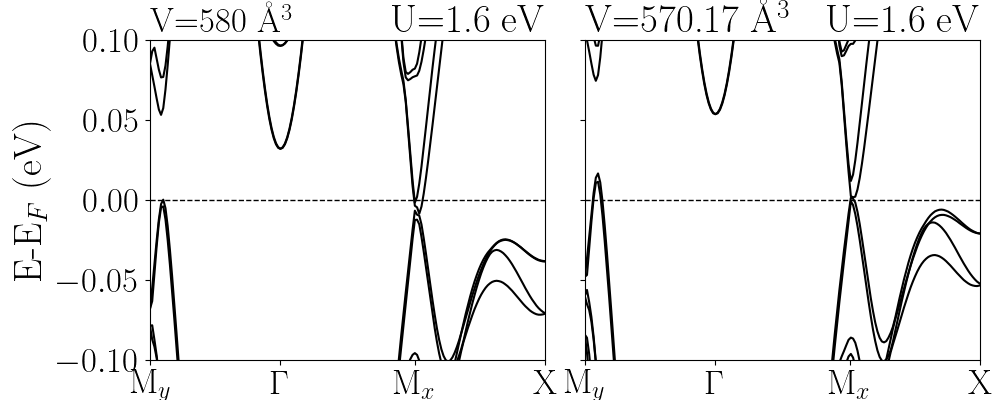}}

\caption{PS: band structure for U =0, 0.5,1.4 and 1.6 eV considering positions (left panel) and volume (right panel) degrees of freedom.}
\label{PS-bands}
\end{figure}

\begin{figure}[H]
\centering
{\includegraphics[width=0.4\textwidth]{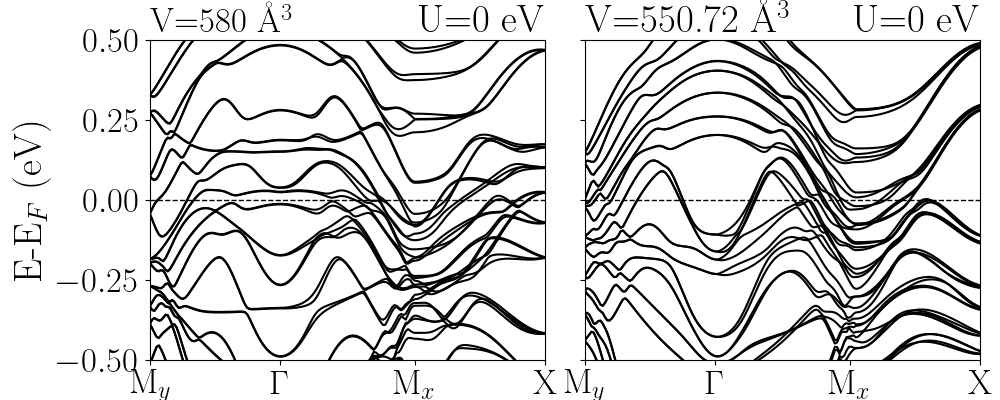}}
{\includegraphics[width=0.4\textwidth]{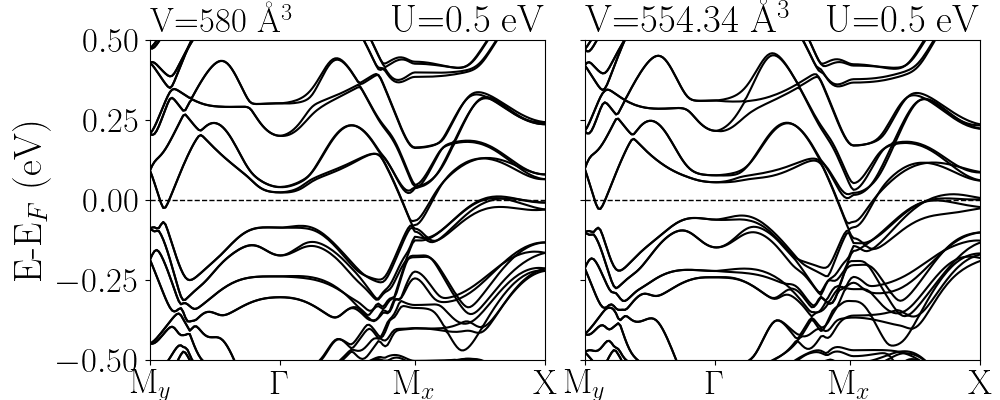}}
{\includegraphics[width=0.4\textwidth]{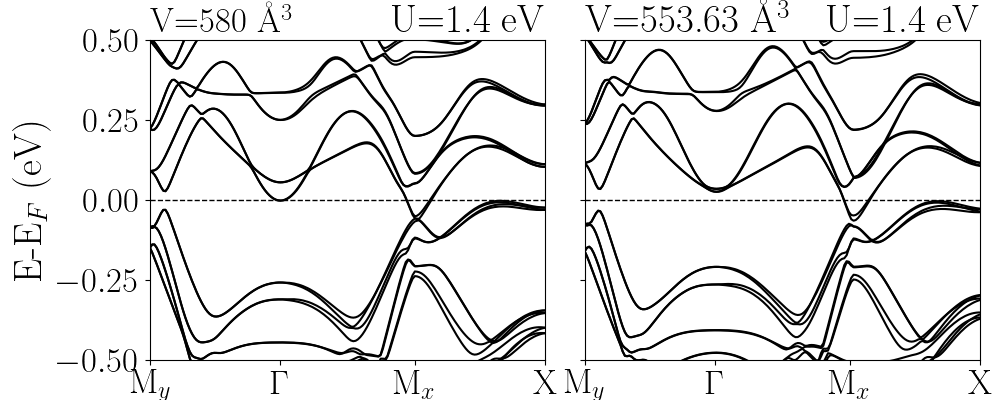}}
{\includegraphics[width=0.37\textwidth]{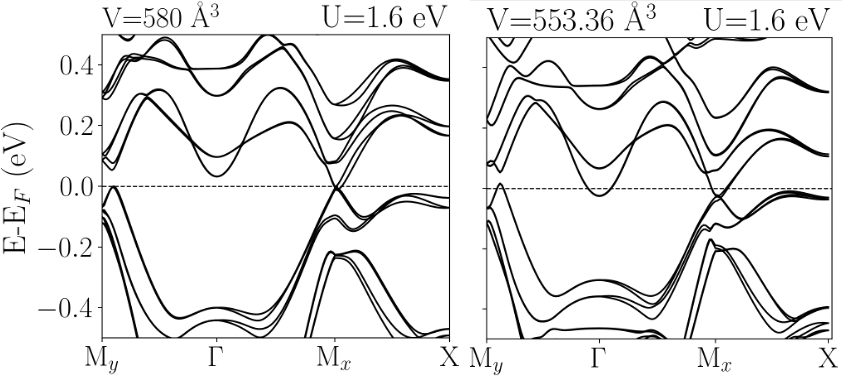}}
\caption{LDA: band structure for U =0, 0.5, 1.4 and 1.6 eV considering positions (left panel) and volume (right panel) degrees of freedom.}
\label{2LDA-bands}
\end{figure}

Fig. \ref{PS-bands} and  \ref{2LDA-bands} shows the band structure of CRO in the Bb2$_{1}$m structure (with the position reported by the experiment at 8 K\cite{yoshida2005}) for PS and LDA approximation, respectively, considering positions (left panel) and volume (right panel) degrees of freedom in the calculation. The calculations were performed for selected U values (U = 0, 0.5, 1.3, 1.4 and 1.6 eV).

\section{Dudarev and Liechtenstein approach.}

\begin{figure}[ht]
\centering
\subfigure[]{\includegraphics[width=0.35\textwidth]{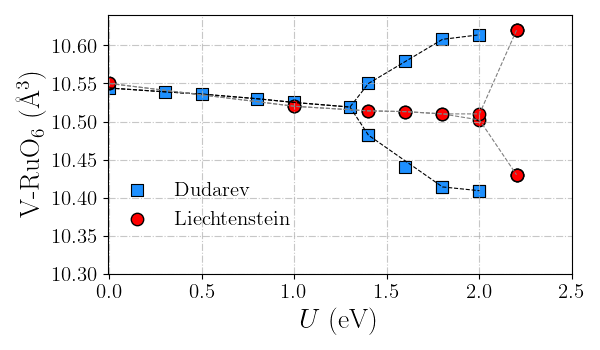}}
\subfigure[]{\includegraphics[width=0.35\textwidth]{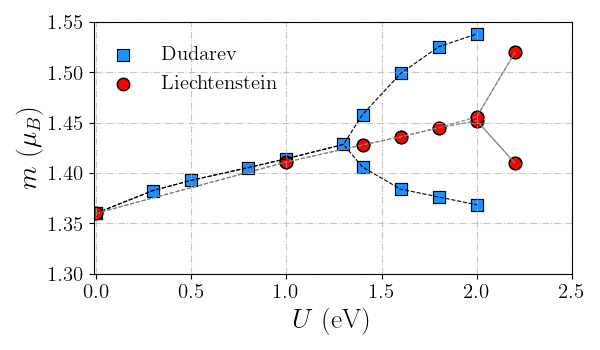}}
\caption{(a) Volume RuO$_{6}$ octahedra (V-RuO$_{6}$) and (b) magnetic moment by Ru atom ($m$) inside of WS radius, for two approximations to include U-Hubbard repulsion: Dudarev and Liechtenstein. Calculation considering positions degrees of freedom along PBEsol approximation. 
\label{li1}}
\end{figure}

Fig. \ref{li1} shows the octahedra volume and magnetic moment by Ru atoms as U-Hubbard increases, along two approximations Dudarev and  Liechtenstein approaches. Here, we can observe that the electronic changes associated with the phase transition appear independent of the method, happening U$\geq$ 1.4 and U$>$ 2.0 for Dudarev and  Liechtenstein, respectively.

\end{document}